# Expanded stability of layered SnSe-PbSe alloys and evidence of displacive phase transformation from rocksalt in heteroepitaxial thin films

Pooja D. Reddy[1*], Leland Nordin[1], Lillian Hughes[2], Anna-Katharina Preidl[1], Kunal Mukherjee[1†]

[1]Department of Materials Science and Engineering, Stanford University, Stanford, California 94306, USA
[2]Materials Department, University of California, Santa Barbara, California 93106, USA

## Abstract

Bulk PbSnSe has a two-phase region, or miscibility gap, as the crystal changes from a Van der Waals-bonded orthorhombic 2D layered structure in SnSe-rich compositions to the related 3D-bonded rocksalt structure in PbSe-rich compositions. This structural transition drives a large contrast in the electrical, optical, and thermal properties. We realize low temperature direct growth of epitaxial PbSnSe thin films on GaAs via molecular beam epitaxy using an *in situ* PbSe surface treatment and show a significantly reduced two-phase region by stabilizing the *Pnma* layered structure out to $Pb_{0.45}Sn_{0.55}Se$, beyond the bulk-limit around $Pb_{0.25}Sn_{0.75}Se$ at low temperatures. Pushing further, we directly access metastable two-phase films of layered and rocksalt grains that are nearly identical in composition around $Pb_{0.50}Sn_{0.50}Se$ and entirely circumvent the miscibility gap. We present microstructural and compositional evidence for an incomplete displacive transformation from a rocksalt to layered structure in these films, which we speculate occurs during the sample cool down to room temperature after synthesis. *In situ* temperature-cycling experiments on a $Pb_{0.58}Sn_{0.42}Se$ rocksalt film reproduce characteristic attributes of a displacive transition and show a modulation in electronic properties. We find well-defined orientation relationships between the phases formed and reveal unconventional strain relief mechanisms involved in the crystal structure transformation, using transmission electron microscopy. Overall, our work adds a scalable thin film integration route to harnessing the dramatic contrast in material properties in PbSnSe across a potentially ultrafast structural transition.

* poojadr@stanford.edu
† kunalm@stanford.edu

## Introduction

The IV-VI semiconductor SnSe-PbSe alloy system hosts a wide array of properties relevant to thermoelectrics,[1–3] infrared detectors and optical modulators,[4–7] ultrathin in-plane ferroelectrics,[8–11] and most recently quantum information sciences via spin qubits and as topological crystalline insulators.[12,13] Long known for its unusual bonding[14,15] and complex lattice dynamics,[16] this alloy system offers the opportunity to study and harness properties at the intersection of structural and electronic phase transitions. Specifically, bulk $Pb_{1-x}Sn_xSe$ (PbSnSe) hosts two closely related structural phases of layered orthorhombic *Pnma* (<450 °C) and *Cmcm* (>450 °C) symmetry from SnSe to approximately $Pb_{0.25}Sn_{0.75}Se$, which have an indirect and direct bandgap respectively, and are trivial insulators.[17,18] These layered phase unit cells are made up of two bilayers (four Sn-Se atomic layers in total) with characteristic A-B stacking. On the Pb-rich side, the bulk stable phase is $Fm\bar{3}m$ rocksalt from PbSe to approximately $Pb_{0.57}Sn_{0.43}Se$;[18] this phase has a direct bandgap and becomes topologically nontrivial for metastable alloys with Sn content increasing beyond $Pb_{0.77}Sn_{0.23}Se$.[12] A miscibility gap or two-phase region is reported for intermediate compositions and the extent of this gap is calculated to increase at lower temperatures.[18–20]

Thanks to the chemical similarity of the Pb and Sn atoms, the layered and rocksalt phases are structurally very closely related to each other. Small, sub-unit cell displacements of atom positions convert one to the other. Indeed, the layered structure is often referred to as a distorted rocksalt structure.[21] Yet, there arise large property differences between these two phases because the 2D-bonded layered phase has ionic and covalent character bonding within layers and Van der Waals (VdW) bonding between layers, while the 3D-bonded cubic rocksalt phase has mixed metallic, ionic, and covalent character bonds in all directions.[22] This unusual combination of large property contrast between the phases while retaining close proximity in structure has the potential for important phase-change devices, if the miscibility gap can be avoided. Recently, results from Katase *et al.* and Nishimura *et al.,* in bulk crystals and reactive solid-phase epitaxy thin films with compositions close to $Pb_{0.50}Sn_{0.50}Se$, show a direct switch between these two crystal structures with no change in composition.[17,23] The altered bonding motifs across the rocksalt to layered-orthorhombic phase transition yield large changes in electronic and thermal properties, accessed by simply varying the sample temperature.[17] Bulk crystals and films were prepared in the rocksalt phase at high temperatures (in excess of 600 °C) and quenched, presumably to avoid conventional diffusive phase transformations to Pb-rich and Sn-rich domains. Upon quenching rapidly to room temperature, deep into the bulk two-phase region of the phase diagram, 87% volume fraction of the sample converted to the layered-orthorhombic phase.[17] The volume fraction increases to 90% at cryogenic temperatures of −170 °C .[17] Heating the sample back up reverts all the layered-orthorhombic phase fraction back to rocksalt. Phase transitions during quenching or at ambient and cryogenic temperatures cannot involve long range diffusion of atoms, and hence are likely to be diffusionless. These transitions could be ultrafast displacive phase transformations without compositional changes, which involve an organized small-scale shuffle of atoms between the related crystal structures.[24] At the same time, other researchers also note the energetic proximity of the two crystal structures and have proposed displacively switching SnSe from a layered to a rocksalt structure by application of intense light-fields by harnessing differences in the real part of the dielectric response.[25,26]

Building off these recent developments, we explore to what extent direct low-temperature synthesis (165–300 °C) of epitaxial PbSnSe thin films controls the final crystal structure. This low-temperature growth is especially important as the bulk phase diagram is poorly characterized at these temperatures. In addition, investigating epitaxial thin films by molecular beam epitaxy (MBE) on GaAs also presents a technologically relevant method to harness the rare electronic and optical properties of both the rocksalt and layered phases in heterostructures, potentially even the displacive transformation, and merge them with mature optoelectronic technologies. In this work, we show that single-phase layered samples can be stabilized deep into the bulk two-phase region on the Sn-rich side by direct low temperature synthesis of epitaxial films on GaAs(001) substrates with binary PbSe buffer layers. Using this low temperature synthesis method, we also prepare two-phase samples at compositions near $Pb_{0.50}Sn_{0.50}Se$ and find microstructural evidence of a displacive phase transformation, which is accompanied by a strong contrast in electronic properties. Additionally, these two phases have well-defined orientation relationships and novel strain-relief mechanisms that allow for minimal strain during the transformation between a 3D and 2D-bonded structure.

## Optimizing SnSe epitaxy on GaAs

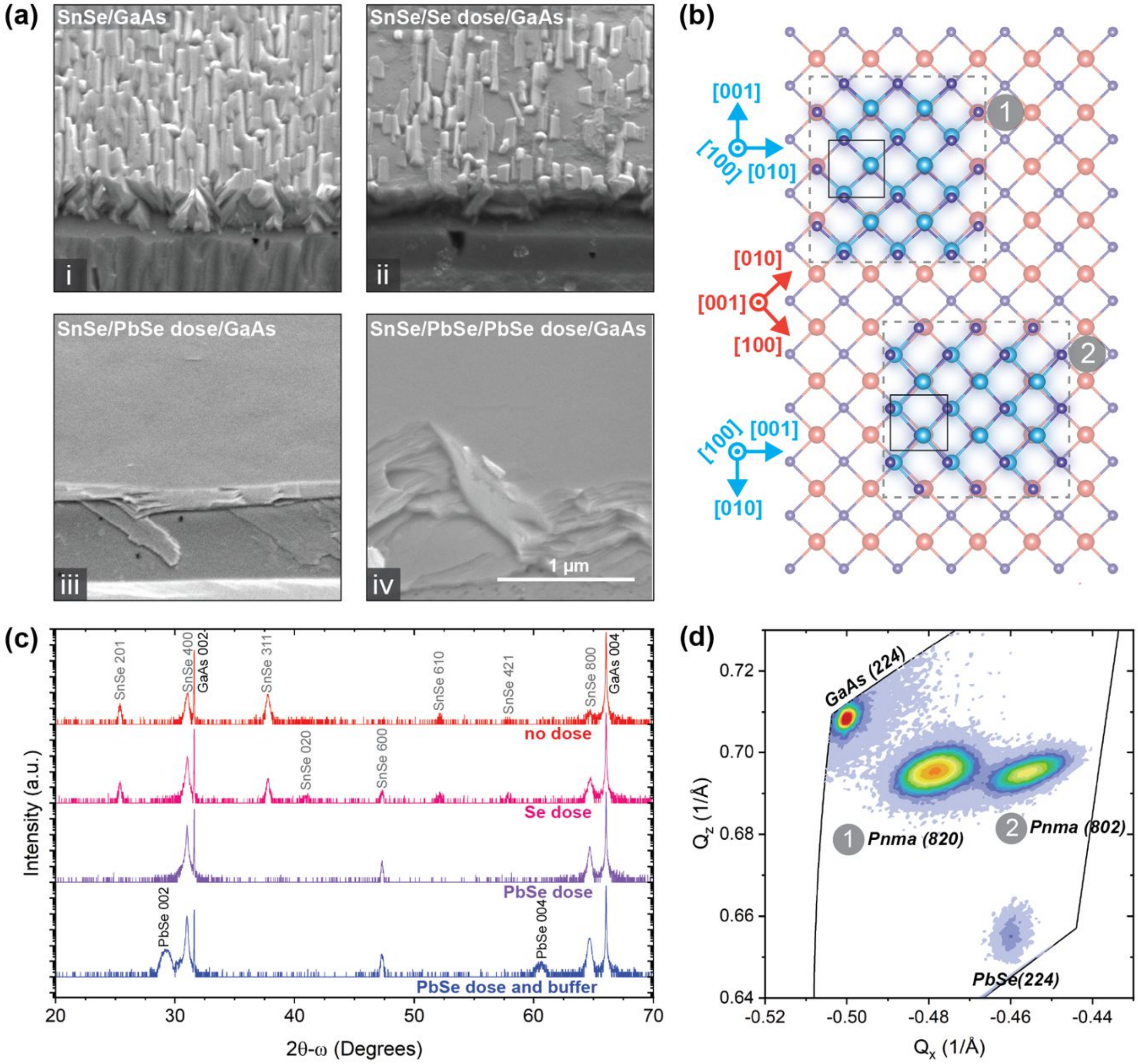

**Figure 1.** (a) SEM images of the cleaved edge of 45° mounted SnSe samples grown with various surface preparation steps: (i) direct growth on oxide-desorbed GaAs substrate, (ii) oxide desorb and Se dose step, (iii) oxide desorb and PbSe dose step, (iv) oxide desorb, PbSe dose, and PbSe buffer steps. (b) Schematic of how the two orientations of SnSe grow on PbSe. Black rectangles show the unit cell of *Pnma* SnSe. (c) Triple-axis symmetric 2θ-ω XRD scans of each of the surface preparations. (d) RSM of SnSe with PbSe dose and buffer surface treatment. *Pnma* SnSe film, PbSe buffer layer, and substrate peaks are labelled.

We first describe epitaxial synthesis of *Pnma* layered SnSe films on GaAs(001) as a precursor to the PbSnSe alloy films. Based on scanning electron microscopy (SEM) images, SnSe nucleated directly on oxide-desorbed GaAs, termed a 'no-dose' surface preparation, results in a rough, faceted surface (Fig. 1ai). Related, a symmetric 2θ-ω X-ray diffraction (XRD) scan reveals multiple out-of-plane (OP) orientations (Fig. 1c) in this sample, and we find the dominant orientations are indexed as (400), (311), and (201). Note that we label the a-axis as across VdW-bonded layers and the b- and c-axes being the zigzag and armchair directions, respectively. As the terminated non-(100) surfaces of SnSe have strong ionic and covalent bonding character, we suspect that these orientations arise because of partial lattice-matching to the GaAs surface. One of the edges of the intersection of the SnSe unit cell and the (311) plane is lattice matched with GaAs, which explains the prevalence of this orientation. (201)-oriented grains grow such that the [010] direction is along the [110] direction of GaAs, which is also fairly well lattice matched. These lattice-matching constraints are relaxed for the largely VdW-bonded (100) surface of SnSe and hence may be preferentially nucleated by growth on a low-energy substrate surface.

Prior work shows a group-VI flux of Se and Te can prepare a GaAs substrate surface and improve the quality of selenide and telluride epitaxial films,[27,28] perhaps because exposure to a chalcogen beam converts the surface layers to be VdW GaSe or $Ga_2Se_3$-like and facilitates VdW-epitaxy.[27,29] Therefore, we expose the GaAs substrate after the oxide desorption step to a Se dose step: Se flux of $2\times10^{-7}$ Torr at 400 °C for 30 s. The effect of this Se flux treatment on the resulting SnSe film can be seen in Figure 1aii. The film has smooth areas, however some faceting is still present. The corresponding XRD data of this Se dose sample mirrors this apparent improvement in epitaxial film quality (Fig. 1c), where the intensity of alternate orientations has decreased compared to the {100} family of planes.

While the Se flux improves SnSe film quality, we have previously shown that a PbSe dose and a thin rocksalt PbSe buffer layer deposited before the growth of PbSe improves structural quality.[30] We extended this method to the growth of SnSe since, as discussed, the structure of PbSe and SnSe are so similar to each other. For the dose, the substrate temperature is at 400 °C so that it is very unlikely that any PbSe molecules adsorb, and the substrate is exposed to a PbSe flux for 30 s. A buffer layer consisting of 10 nm of PbSe is then grown at 330 °C. Both the SnSe film with a PbSe dose, and the sample with a PbSe dose and buffer layer produced smooth films (Fig. 1aiii,iv). The symmetric $2\theta$-$\omega$ scans of these samples show only {100} out-of-plane (OP) oriented SnSe and a complete elimination of alternatively oriented crystallites. The nonpolar (001) surface and nondirectional bonding nature of PbSe seems to facilitate ordered growth of the (100) oriented SnSe on GaAs (001). These structural improvements lead to an order of magnitude higher carrier mobility (Table S1). And so, by exploring various substrate preparation steps, we can achieve the growth of smooth epitaxial thin films of orthorhombic SnSe on GaAs (001).

Given that the (100) surface of the SnSe unit cell is (slightly) rectangular, while that of (001) PbSe and (001) unreconstructed-GaAs is square, it is worth considering the in-plane (IP) orientation of the SnSe epilayer. Figure 1d shows a reciprocal space map (RSM) around the (224) reflection of the GaAs substrate, with the 224 reflection from the PbSe buffer layer, and two reflections for the SnSe film. We index the SnSe reflections as (820) and (802). Thus, on PbSe/GaAs, the rectangular base of SnSe orients in two ways, perpendicular to each other. Figure 1b shows a schematic of these two orientations of SnSe on PbSe, with the unit cells of SnSe marked. Because of the two perpendicular orientations, we can obtain information on both IP lattice constants of the *Pnma* phase from one RSM. Once corrected for the structure factor of the two reflections, we find approximately equal volume fractions of each of the two grain orientations.

We note that the IP lattice parameters of our SnSe film deviate from that of the bulk. While there is some discrepancy on the lattice constants of bulk SnSe, using the most recent values obtained by Nishimura *et al.*[17] we find our 'b' parameter along the zig-zag axis is nominally under tension by 0.7%, while the 'c' parameter along the armchair direction is compressed by 1.3%. The unconstrained OP 'a' parameter is extended by 0.13%. Much of this elastic strain is in qualitative agreement with the effects of thermal expansion mismatch between the large and anisotropic thermal expansion coefficients of SnSe with respect to GaAs during cool down from growth temperature to room temperature. The arm-chair axis notably has a negative thermal expansion coefficient so we expect the unit cell is compressed in this direction upon cooling down if locked to the substrate, while the zigzag direction has a positive thermal expansion coefficient and hence is strained in tension.[31] We will evaluate these strain effects further in a separate study.

## Extension to single-phase layered PbSnSe alloys

Continuing to use the PbSe dose and thin PbSe buffer layer to preserve the epitaxial nature of the thin films, we obtain single-phase layered films up to $Pb_{0.45}Sn_{0.55}Se$ at growth temperatures of 165–300 °C. We use XRD to estimate the composition of our samples in conjunction with published trends of the OP lattice constant of bulk PbSnSe as a function of composition.[17] As previously mentioned, the OP lattice constant of the thin film SnSe sample is strained 0.13% compared to the bulk. Since we expect the thermal expansion coefficient of PbSnSe alloys to not vary greatly with composition,[31–33] we assume a constant value of 0.13% residual strain in the thin film samples across composition. Note that the calculated compositions of our samples shift by only 2-3% if not accounting for this strain. Additionally, X-ray photoelectron spectroscopy (XPS) results show that the alloy composition is constant throughout the epilayer thickness (Fig. S1). Figure 2 shows all films produced in this study (also Table S2). This includes all single-phase layered samples, which extend beyond the $X_{Sn}$≈0.8 solubility limit at 300 °C in the bulk phase diagram (Fig. 2, black lines). We can confirm the layered alloys grown at these low growth temperatures are the *Pnma* orthorhombic phase as expected and not the high-temperature *Cmcm* phase by detecting an (811) peak via RSM, which is forbidden for the *Cmcm* phase (S2, Fig. S2). The layered structure of these alloy samples is corroborated in the inset transmission electron microscopy (TEM) image in Figure 3e of a $Pb_{0.29}Sn_{0.71}Se$ film which also shows epitaxial registry of atoms at the interface. The wider view scanning transmission electron microscopy (STEM) image shows the thin PbSe buffer layer on the GaAs substrate, on top of which is the PbSnSe alloy film (Fig. 3e). We note threading dislocations through the PbSnSe alloy film, originating from the PbSe/PbSnSe interface but see no signs of phase separation.

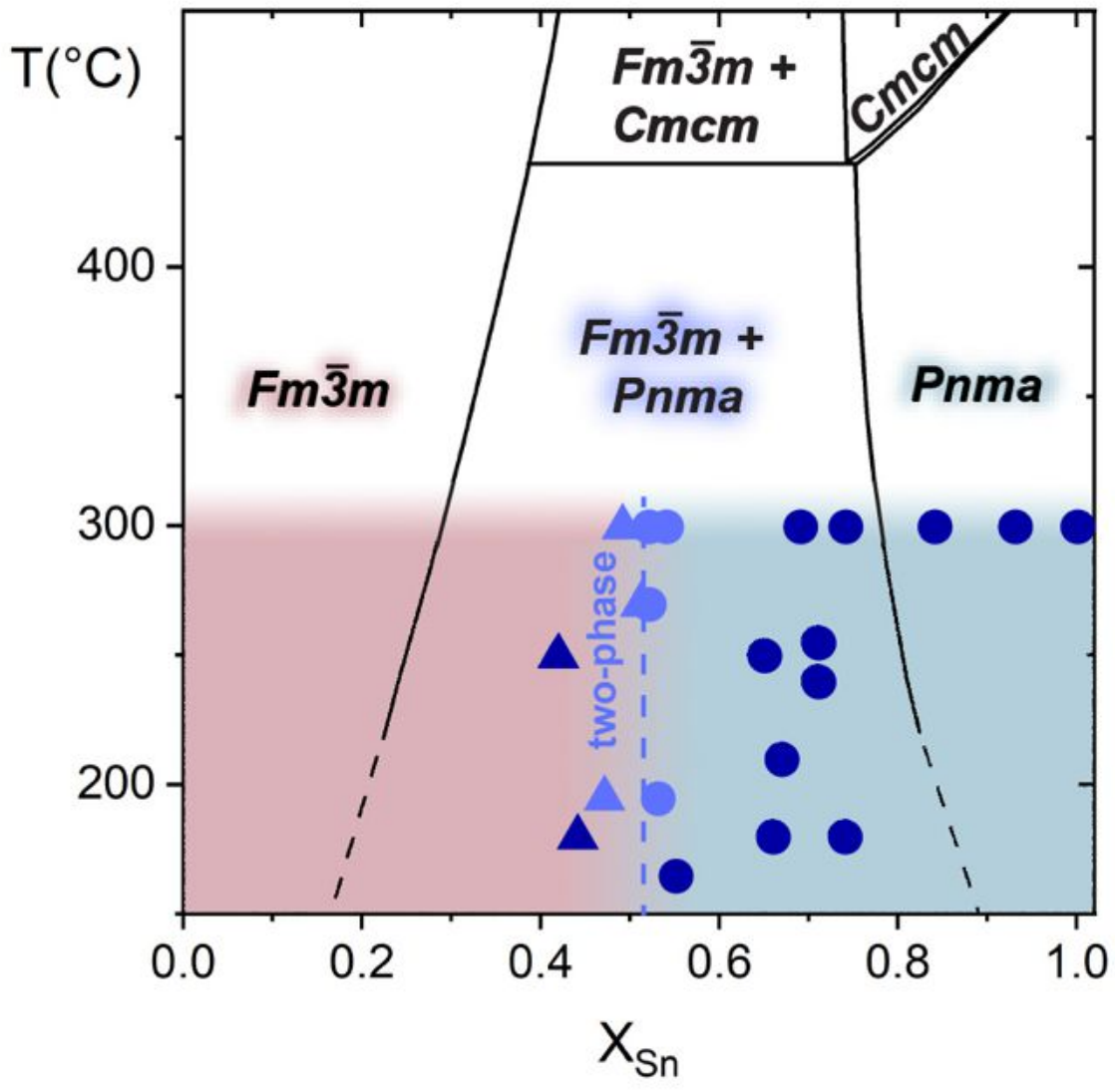

**Figure 2.** Black lines mark the phase diagram of the bulk PbSe-SnSe system, adapted from reference 19. Growth conditions of MBE samples mentioned in this work are overlaid. Layered phases are marked by circles and cubic rocksalt phases are marked by triangles. Single-phase samples are marked in dark purple, while the approximate compositions of each phase in two-phase samples are marked in light purple. Shaded color shows the metastable phases accessed with MBE grown thin films; blue on the right is for layered phase, red on the left is for rocksalt phase. The dashed purple line denotes the two-phase displacive transformation region at growth temperatures below 300 °C.

The largest set of layered phase samples in this study are grown at 300 °C and provide the most consistent basis with which to probe trends in the unit cell dimensions with alloy composition. Figure 3a shows the general sample structure. As the composition of Sn decreases in the alloy, the (800) *Pnma* reflection position moves to smaller angles indicating the OP lattice constant increases (Fig. 3d). The full width at half maximum (FWHM) of the double-axis rocking curves of the (800) film peak reflections range from 1400–1770 arcseconds (Fig. S3), on par with previously grown thick orthorhombic SnSe samples.[34] Experimental quantification of PbSnSe unit cell dimensions as a function of Sn composition is presented in Figure 3b. As before, these are obtained from a single RSM scan around the (224) GaAs peak as the 'b' and 'c' axes are simultaneously accessible due to the two grain orientations of *Pnma* PbSnSe on PbSe/GaAs. The increase of the OP 'a' lattice constant, and the approach of the 'b' and 'c' axes with decreasing Sn content is consistent with an approach towards the cubic rocksalt phase. Figure 3c shows how the unit cell volume of PbSnSe thin films change with composition for the layered *Pnma* phase, along with values based on the lattice constant trend of rocksalt $Fm\overline{3}m$ PbSnSe thin films synthesized under similar conditions.[35] With important implications for the subsequent discussion on phase transformation, the extrapolated unit cell volume change between the layered and the rocksalt phase is at most 5 $Å^3$ in our thin films, with the difference decreasing as Sn composition decrease. In other displacive phase transformations such as austenite to martensite[36] or for shape memory alloys,[37] the volume change between phases is about 20–25 $Å^3$. With a smaller volume change, the phase transition from layered to rocksalt in PbSnSe thin films devices will require less energy. Moreover, the phase transition should impart smaller transformation-strains and thus the material will experience less fatigue.

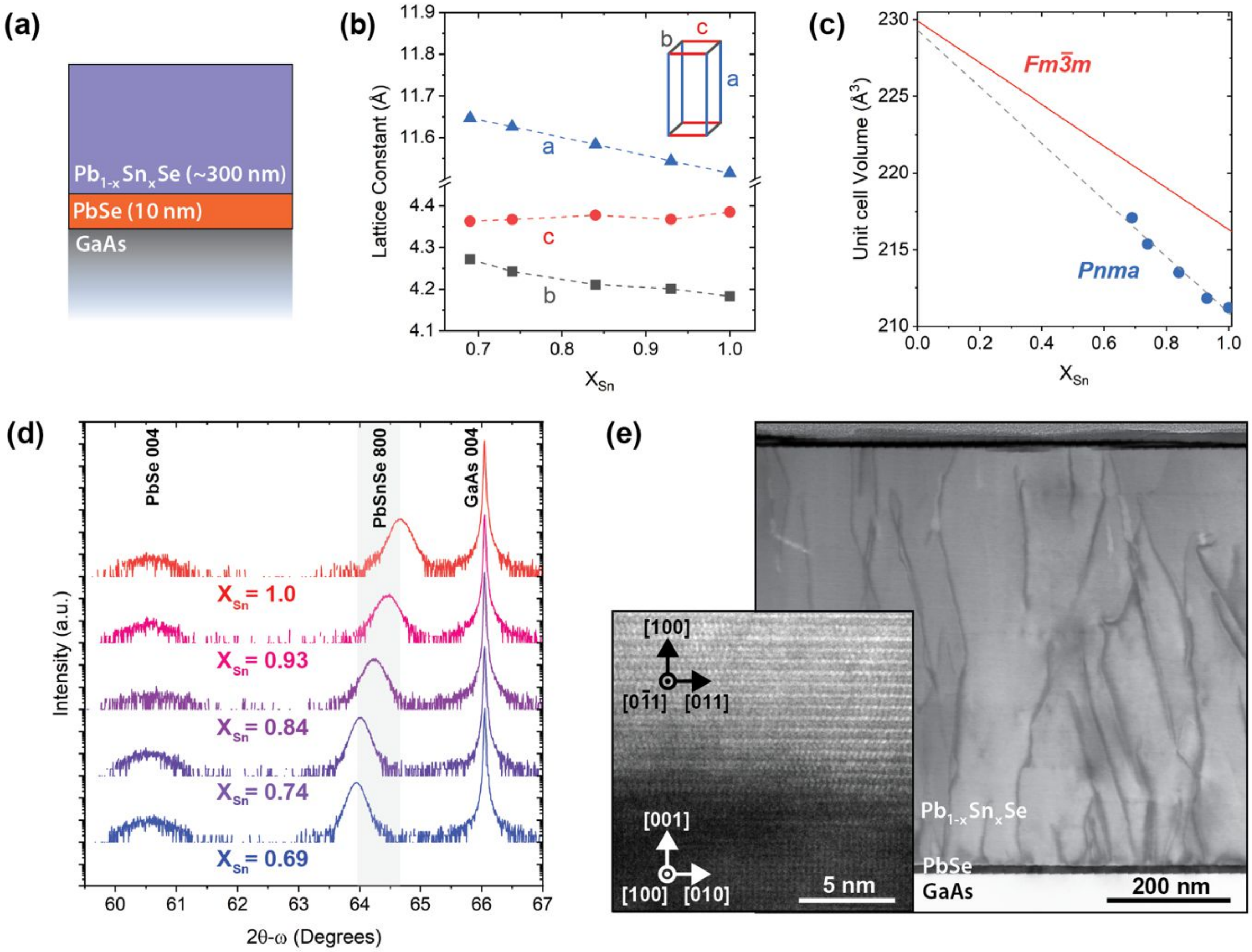

**Figure 3.** (a) Sample structure of thin film *Pnma* PbSnSe alloys grown at 300 °C. (b) Lattice constants and (c) unit cell volume as a function of Sn composition for these layered phase films. Unit cell volume of Pb-rich $Fm\bar{3}m$ rocksalt phase PbSnSe adapted from reference 35. (d) Triple axis symmetric 2θ-ω scans, where the alloy compositions are determined from the a-axis lattice constant data in Figure 3b. (e) Cross-sectional STEM image of a layered phase $Pb_{0.29}Sn_{0.71}Se$ film, including the PbSe buffer layer and GaAs substrate. Inset shows higher resolution TEM image of PbSnSe/PbSe interface with respective lattice directions marked.

Further reducing the film composition to nominally $Pb_{0.50}Sn_{0.50}Se$ results in two-phase mixtures of layered and rocksalt grains. We will discuss these two-phase samples shortly. Interestingly, reducing the growth temperature down to 165 °C does not alter our ability to push the single-phase region on the layered-orthorhombic side beyond this limit. Preliminary experiments of growth at temperatures of 150 °C yielded signs of polycrystalline nucleation based on reflection high-energy electron diffraction (RHEED) images (Fig. S4), and hence we do not extend to temperatures below this.

**Two-phase layered and rocksalt films and evidence of a displacive phase transformation**

Close to a nominal composition of $Pb_{0.50}Sn_{0.50}Se$, we synthesized multiple PbSnSe films that show the co-existence of the layered and rocksalt phases. Figure 4 shows an RSM of a two-phase film grown at 300 °C. There is a (224) peak corresponding to the rocksalt phase of PbSnSe as expected. In addition, we find a single peak corresponding to the layered phase instead of the two distinct (820) and (802) peaks seen in Section 2.2. This single peak has a different OP spacing ($1/Q_z$) but the same IP spacing ($1/Q_x$) as the rocksalt phase, suggesting the potential of a coherent or lattice-matched relationship between the two phases. This type of RSM is typical for all the two-phase samples synthesized. Based on the lattice constant trend in Figure 3b, we expect the IP lattice constants of the orthorhombic unit cell to coincide around $Pb_{0.50}Sn_{0.50}Se$. It seems that in these two-phase samples, the Pb composition of the layered-orthorhombic phase is high enough such that the previously distinguishable (802) and (820) reflections have approached each other. We consider whether this layered phase could also be the high temperature *Cmcm* phase, since previous work has shown that adding Pb depresses the phase transition temperature between the *Pnma* and *Cmcm* phases.[38] Yet once again, we detect a symmetry-forbidden peak for the *Cmcm* phase via RSM analysis, thus confirming our assignment of *Pnma* (S2, Fig. S2). We estimate the composition of the layered and rocksalt phases separately using the lattice parameters from the RSM, employing the same procedure used for the single-phase samples. Remarkably, we find that the rocksalt phase and the layered phase have very similar compositions (shown in Fig. 2) at around $Pb_{0.50}Sn_{0.50}Se$. This contrasts with what may be expected in classical diffusive phase separation where compositions are at the boundaries of the two-phase region.

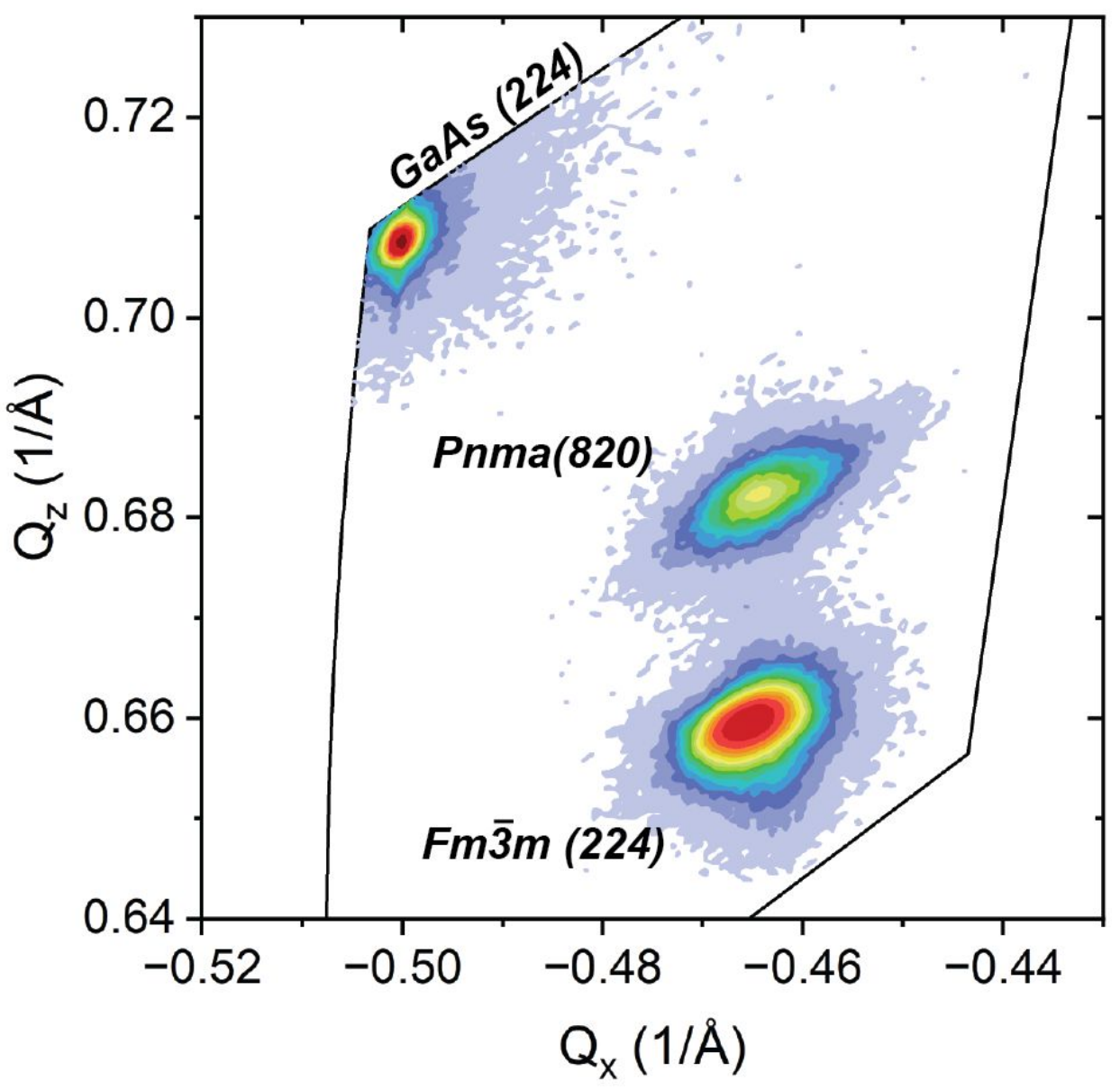

**Figure 4.** RSM of a two-phase PbSnSe alloy grown at 300 °C where the layered and rocksalt phases have compositions of approximately $X_{Sn}$=0.52 and $X_{Sn}$=0.49, respectively. Substrate and film peaks labelled. The PbSe buffer layer peak is masked by the film's rocksalt peak.

Raman scattering measurements further corroborate the Sn content of the layered phase in the two-phase samples. Figure 5a shows the room-temperature Raman spectra for samples grown at 300 °C, first showing the single-phase samples as a guide (Fig. 5ai) and then the two-phase samples (Fig. 5aii). Looking at the single-phase samples, the SnSe sample has five Raman mode peaks, which are characteristic of bulk *Pnma* SnSe: $A_{g(2)}$, $B_{3g(2)}$, $A_{g(3)}$, $B_{1g(2)}$, $A_{g(4)}$, where the $B_{1g(2)}$ mode is hidden in the background for pure SnSe. As the Pb composition increases, the $A_{g(2)}$ and $B_{3g(2)}$ clearly shift to lower wavenumbers (soften). Figure 5b shows the position of the $A_{g(2)}$ mode peak as a function of alloy composition. The $A_{g(2)}$ mode corresponds to the buckling of the bilayers by moving Sn/Pb and Se atoms from the same atomic layer in opposite directions along the tall 'a' axis, and does not directly stretch or compress Pb/Sn-Se bonds. Thus the softening is based on the alloying of a heavier Pb atom which increases the reduced mass of the system.[41] The intensity of the Raman peaks in the two-phase sample (Fig. 5aii) are lower but the peak signature resembles that of the layered phase. We propose that the weaker intensity arises since the sample also consists of the rocksalt phase where first order Raman scattering is forbidden.[42,43] The $A_{g(2)}$ mode of the layered phase in the two-phase samples is clearly resolved, and the peak position follows the (extrapolated) trend with composition of the single-phase samples (Fig. 5b). Therefore, XRD-calculated composition and Raman scattering indicate that we indeed have two-phase PbSnSe alloys with rocksalt and layered phases that are the same composition, as would be expected when a displacive phase transformation occurs. We further corroborate the lack of compositional contrast between phases via SEM energy dispersive X-ray spectroscopy (SEM-EDS) (Fig. S5), and STEM energy dispersive X-ray spectroscopy (STEM-EDS) described subsequently. At this stage, we think it is unlikely that the sample separated into rocksalt and layered phases during growth as this would lead to greater compositional variation due to fast atom diffusion on the growth surface. We speculate that the $Pb_{0.50}Sn_{0.50}Se$ samples are likely in the rocksalt structure at growth temperature, and they undergo a partial phase transformation in the growth chamber during the cooldown step to room temperature to become two-phase.

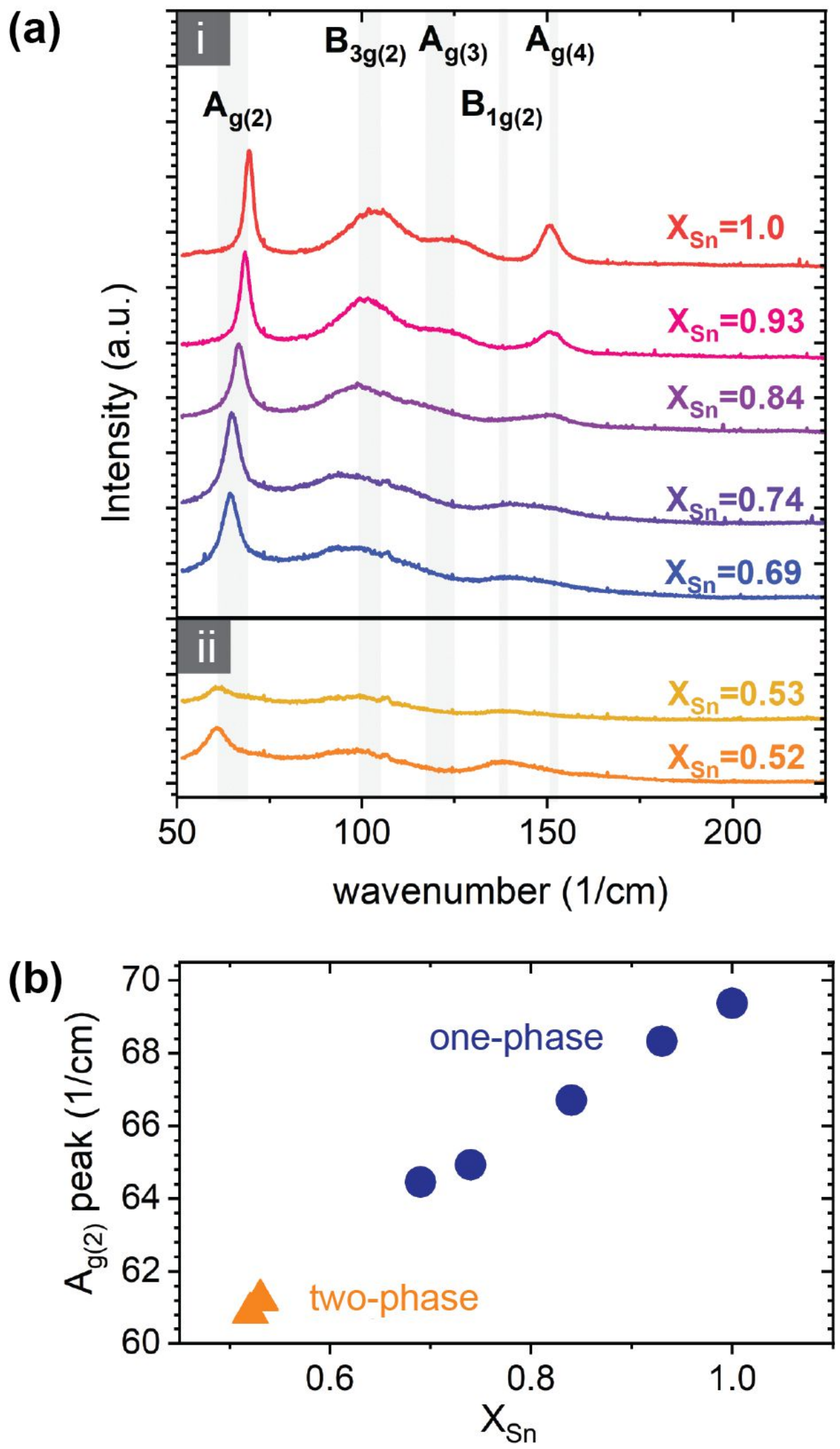

**Figure 5.** a) Raman spectra of samples grown at 300 °C with only the composition of the layered phase marked, including i) single-phase layered samples, and ii) two-phase layered and cubic rocksalt samples. b) Position of the $A_{g(2)}$ peak as a function of layered phase composition.

**Microstructure of the 3D to 2D-bonded displacive phase transformation**

Rocksalt/layered interfaces are a unique case where the bonding motif changes from 3D to 2D. We look at the microstructure of an as-grown two-phase sample to understand how these domains co-exist, and find remarkable mechanisms involved in these crystal structure transformations. Figure 6a shows a dark field (DF) STEM image of a two-phase film on a GaAs substrate with a PbSe buffer layer. Based on XRD data, this sample is made up of 40% rocksalt and 60% layered grains. The linear features in the film are not dislocations seen previously for SnSe, but rather denote interfaces or grain boundaries. The PbSnSe samples may look columnar at first glance, but a closer inspection reveals that the columns are neither continuous nor perfectly vertical and do not originate at the substrate interface as might be expected for a film that is phase separating under growth conditions.[44–47] This supports the assertion that two-phase samples around $Pb_{0.50}Sn_{0.50}Se$ are rocksalt during growth and undergo a partial phase transformation to the layered phase during cooldown to room temperature.

Before we discuss the atomic structure of the interfaces, we reiterate that XRD shows that the IP spacing (in both directions) is the same for both phases at 6.06 Å (note that the IP axes of the Pnma unit cell is 45° rotated to the rocksalt and the dimensions are matched after multiplying by $\sqrt{2}$). This arises as the unit cell dimensions of the armchair and zigzag directions approach each other with increased Pb alloying. The OP spacing between the phases is nevertheless mismatched by 3.3%; the layered phase is shorter at 11.72 Å compared to a height of 12.12 Å for two rocksalt unit cells. Thus, phase transformation need only accommodate this OP mismatch and a complete transformation from (001)-oriented rocksalt to (001)-oriented layered phase would result in the entire film shrinking in height by 3.3%, without any lateral strain.

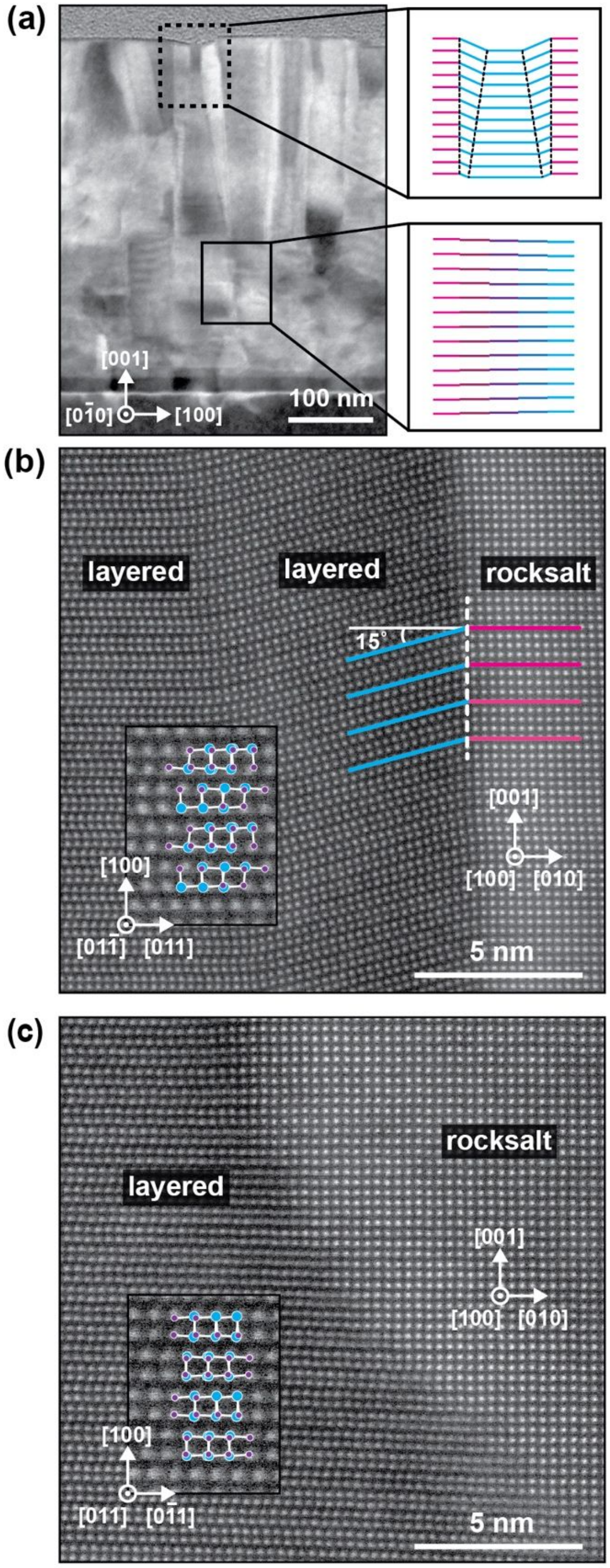

**Figure 6.** STEM images of a PbSnSe alloy grown at 300 °C with layered and rocksalt phase compositions of approximately $X_{Sn}$=0.53 and $X_{Sn}$=0.49, respectively. (a) DF STEM image of alloy, including the lattice directions of the GaAs substrate and PbSe buffer layer. (b) High resolution High-angle annular dark-field (HAADF) STEM image of a vertical grain boundary with an inset of the layered phase orientation. Here select lattice planes (blue and red) and the grain boundary (dashed white) are marked. (c) High resolution HAADF-STEM image of a curved grain boundary with an inset of the layered phase orientation.

We find that the interfaces between phases are parallel to the [100] direction of the rocksalt phase and the <011> directions of the layered phase. Overall, we see two types of interfaces, the first of which we will discuss is distinctly vertical. A typical vertical interface structure seen in the film is highlighted with a dashed box and shown schematically in Figure 6a. We find regions of depressed or sunken (100)-oriented layered phase with angled transition regions (Fig. S6b also). Figure 6b shows the atomic structure of one half of this structure. Looking from right to left, we see a large domain of untransformed rocksalt phase, followed by a lattice-matched transition to a few nm wide region of a layered phase that is tilted 15°. The tilted lattice planes then bend or flex over an extended distance to become horizontal once again, essentially forming a low-angle tilt boundary (Fig. 6b). Across more instances of narrow regions with tilted lattice planes, we note that the tilt angle is in the range of 12- 16° and has variable widths, often largest at the surface and narrowing towards the base of the film. Based on the lattice constants from XRD of this sample, 15° is the tilt angle needed for the orthorhombic layers (11.72 Å spacing) to match to every other unit cell of the rocksalt structure (12.12Å spacing), without using strain. This can be clearly seen via the select {100} family planes marked one unit cell apart for the tilted layered grain and two apart for the rocksalt grain, in Figure 6b. We measure the film is 3% thinner in this transformed region compared to adjacent rocksalt layers, confirming a complete transformation in this domain.

The use of an asymmetric tilt boundary between dissimilar crystals has been theoretically proposed as an alternate mechanism for lattice-matching,[48,49] and our results clearly show this in practice in semiconductors. Remarkably the two phases are only different in the OP spacing and thus a simple shearing transformation along the (001)-rocksalt plane can achieve an unstrained and coherent interface. We speculate that this shearing of the rocksalt layer into the tilted layered phase is achieved by the vertical glide of dislocations. Such orientation relationships achieved by shearing is quite typical of microstructures in displacive transformations.[50] The volume constraint of the phase transformation in a thin film likely persuades the tilted layers to straighten out again via the low-angle tilt boundary. Indeed, layered materials have been shown to host tilt-boundaries that extend over multiple unit cells.[51] Looking closer at the atomic structure across the interface, we also find that the transition from rocksalt to the tilted layered phase is not abrupt but rather occurs over a few unit cells (Fig. 7a). Figure 7b shows the lateral shift, or disregistry, between atoms within what becomes an orthorhombic bilayer. As would be expected, the shift is zero in the rocksalt phase and near zero in the layered phase, with a small lateral shift in the tilted grain that corresponds to the dislocation content of the tilt boundary. Looking between the bilayers, we find the regular A-B stacking order is disrupted temporarily in the tilted segment due to the shearing (Fig. 7a). Figure 7c shows this clearly, where there is zero lateral shift between layers as would be expected in the rocksalt phase. However, in the tilted phase, the lateral shifts for two adjacent bilayers are both positive, thanks to the disregistry incorporated by the dislocation content of the asymmetric tilt boundary. Once the layers bend through the low angle tilt boundary, the disregistry is restored to one bilayer having a positive lateral shift while the adjacent bilayer has a negative lateral shift, as we expect for the A-B stacking layered phase. Figure 7d also shows a smooth transition in the OP lattice spacing between rocksalt and layered, through the intermediate tilted layered phase. Ultimately, the more than 3% OP lattice mismatch between the rocksalt and layered phases is accommodated via the dislocation character of the tilt boundaries.

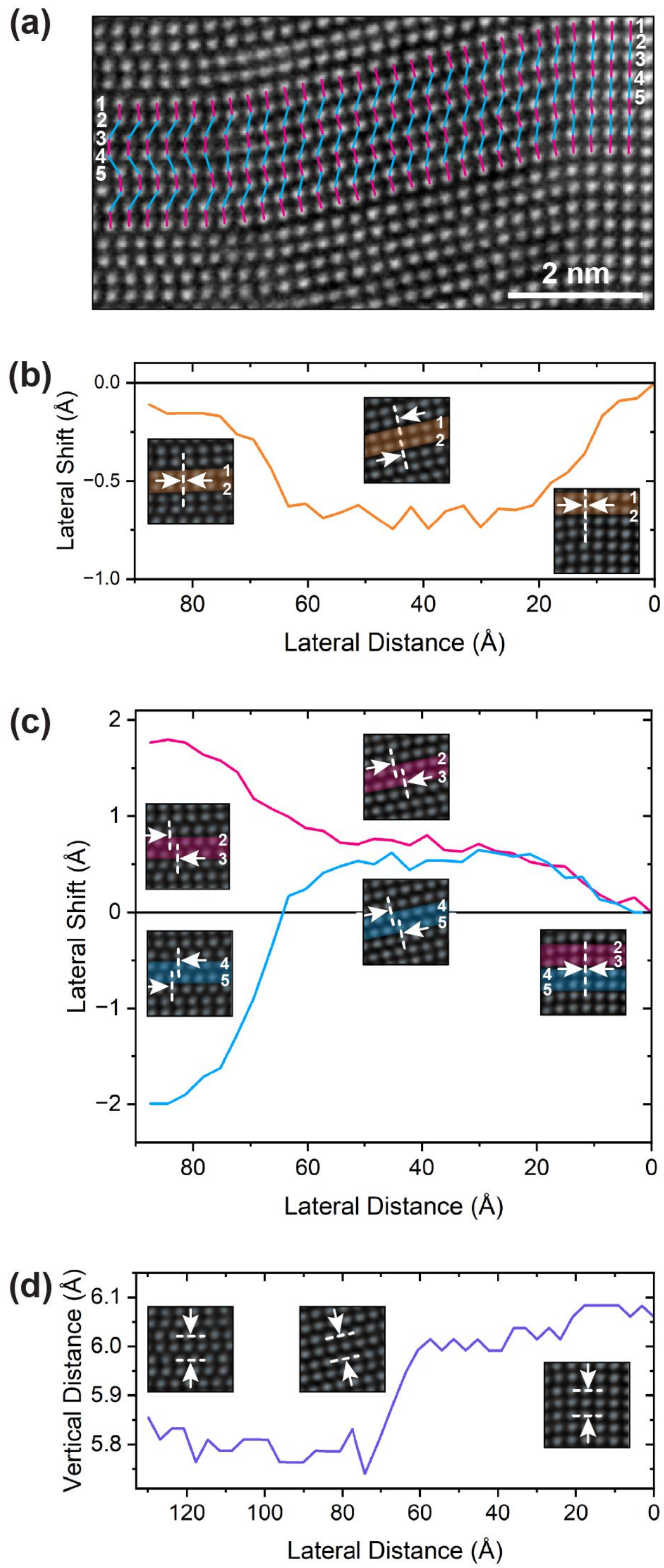

**Figure 7.** (a) A magnified view of the vertical grain boundary in Figure 6b, with lines highlighting lateral shift to make A-B vs A-A' stacking apparent. Lateral shift (b) within a bilayer, and (c) between bilayers of the layered phase as the structure transitions to the rocksalt structure. (d) OP spacing across the vertical grain boundary structure. The half-unit cell height of the layered phase transitions to become one unit cell height of the rocksalt phase.

The solid box in Figure 6a shows an example of the second type of interface that is notably curved and non-vertical. Such interfaces occur in and around smaller volumes of untransformed rocksalt phase and all the lattice planes are horizontal in both phases. Figure 6c has a closer view of the structure. We see the unit cell approximately double in height when going from the rocksalt to layered phase. The atomic columns of every other layer of the rocksalt crystal appear blurred across the interface (Fig. 6c), hinting at a diffuse transformation boundary when viewed in projection. A clearer section of the interface in Figure 8a reveals that the rocksalt structure does indeed smoothly transition to the layered structure over a few unit cells. We resolve this transitional region by tracking the lateral shift, or disregistry, of two adjacent layers, showing a transition from no lateral shift in the rocksalt phase to the A-B bilayer stacking in the layered phase (Fig. 8b). In contrast to the first type of interface, the lateral shift within a bilayer is essentially zero across the interface as there are no tilt-inducing dislocations. Figure 8c shows a smooth transition in the OP lattice spacing between the rocksalt and layered phases. Surprisingly, we see no dislocations in this section of the curved interface mediating the mismatch in OP spacing. We offer three reasons for this absence. First, the gradual transformation of the crystal as opposed to an abrupt interface has the effect of spacing out the dislocations along the vertical direction. Second, the curved or angled nature of the interface reduces the effective mismatch as the phases are IP lattice matched.

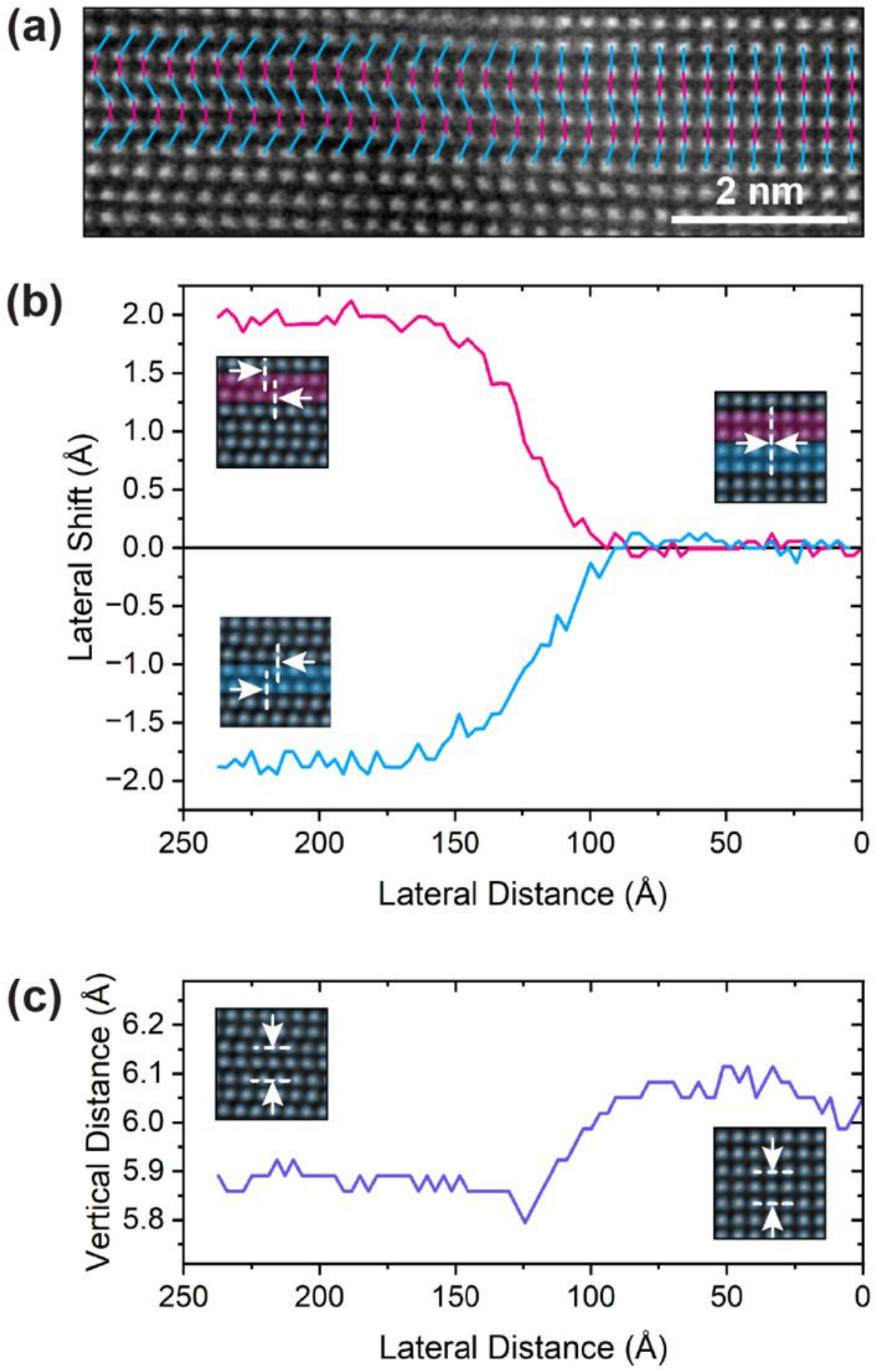

**Figure 8.** (a) A magnified view of the curved grain boundary in Figure 6c, with lines highlighting lateral shift to make the A-B stacking apparent. (b) Lateral shift between two adjacent layers in the layered phase as the structure transitions to the rocksalt structure. (b) OP spacing across the curved grain boundary structure. The half-unit cell height of the layered phase transitions to become one unit cell height of the rocksalt phase.

Third, we note a slight bending of the lattice of the layered phase across the curved interface, indicating another mechanism for accommodating mismatch. These curved interfaces might be able to accommodate more strain elastically, hence the lack of dislocations.

Based on these images, we speculate that phase separation is initiated at points in the rocksalt layer such as a region of roughness on the surface or a grain boundary. Overall, the layered phase prefers to transform to (100)-oriented layered domains in our films, possibly a result of the lower energy in this orientation by exposing the VdW-bonded plane to the surface. At this stage, we do not have sufficient resolution to comment on stacking or rotational faults that arise as the high-symmetry rocksalt transforms to the lower symmetry layered structure, or when domains merge. We do see preliminary evidence of edge dislocations forming between the domains of a layered phase underneath a rocksalt section (Fig. S6a), likely because of the layered phases growing and merging from either side.

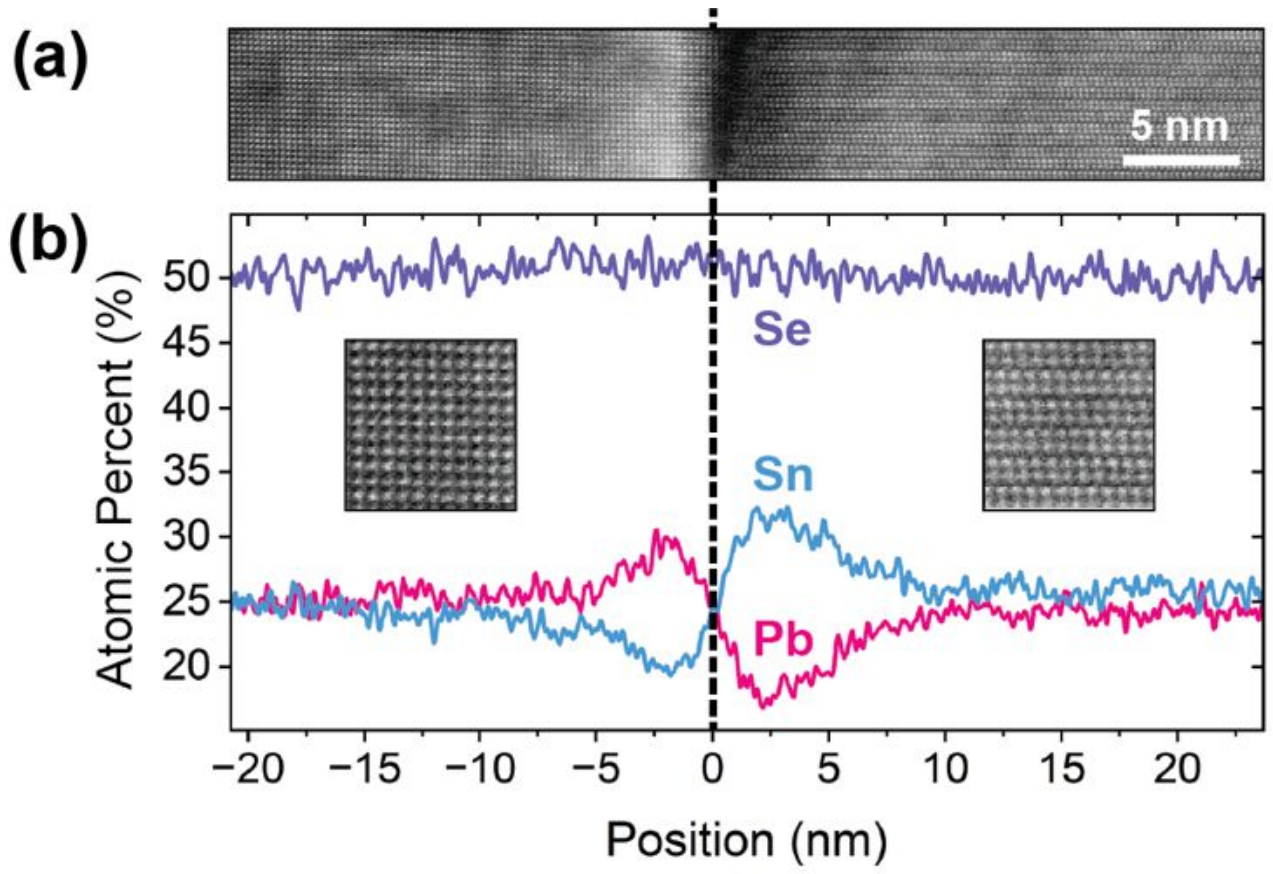

**Figure 9.** (a) HAADF image, which shows the change in z-contrast across a grain boundary between a rocksalt (left) and layered (right) grain. (b) Composition profile in atom percent of Pb, Sn and Se acquired from a STEM-EDS line scan of the same area, showing local diffusion of Pb and Sn at the grain boundary.

Large scale phase separation into Pb-rich rocksalt and Sn-rich layered phases are kinetically hindered at the low temperatures used in this study. Yet, it is useful to ask whether this is also true on the nanoscale at interfaces. Images such as those in Figures 6b and 6c, generated using z-sensitive high-angle annular dark-field (HAADF) contrast, already suggest this is the case. While the interiors of grains have a similar intensity, there is contrast variation near the grain boundaries, with a brighter line of contrast on the rocksalt side, and a darker line of contrast on the layered grain side (Fig. 9a). We quantify composition using standardless STEM-EDS, where the error is typically 2–4%. A line scans over the area shown in Figure 9a show that the Se composition in constant across the boundary at 50 % (atom percent) as expected, and the composition within the bulk of both grains is 25% Sn and 25% Pb (Fig. 9b). This provides independent validation of the XRD composition calculations which estimate that the as-grown two-phase samples have compositions of about $X_{Sn}$=0.5 (Fig. 2). Therefore, EDS allows us to rule out significant composition changes during phase separation and supports our hypothesis of a displacive transformation mechanism. The EDS data also shows that there is indeed local diffusion across the grain boundary. The Sn composition is depleted down to $X_{Sn}$=0.4, 10 nm into the rocksalt phase, and the Sn composition is enriched up to $X_{Sn}$=0.65, 10 nm in the layered phase, trending towards the bulk phase stability limit. Assuming the film is idle at 300 °C for an hour and using a published value of Sn diffusivity in PbSe $D = 2\times10^{-17}$ $cm^2/s$, the characteristic diffusion distance $\sqrt{Dt}$ of 2.5 nm is in reasonable agreement.[52]

This insight into localized composition variation allows us to further formalize a theory for why only a partial displacive transformation occurs in the $Pb_{0.50}Sn_{0.50}Se$ samples. The layered phase likely forms while the sample is still at elevated temperatures, creating grain boundaries. At these temperatures where there is capacity for interdiffusion, group IV atoms are able to diffuse in the absence of a concentration gradient, to create more stable alloys right at grain boundaries (Pb-rich rocksalt phase, Sn-rich layered phase). We do not understand what is pinning the interface in one position long enough for diffusion to occur, but we expect the onset of nanoscale diffusion to make phase switching via interface motion more challenging. Pb-rich rocksalt material will resist switching to a layered structure and vice versa for the Sn-rich layered material. More complete and reversible phase transformation is possible if the sample is single-phase rocksalt at ambient temperature and a displacive transformation is initiated by further cooling the sample. This would reduce the incidence of nanoscale diffusion across the rocksalt/layered grain boundaries, which is likely enabled by elevated temperatures. We will explore the transformation of such a sample next.

In summary, we show that epitaxial PbSnSe samples undergo a displacive transformation aided by close-to-perfect lattice-matching in two IP directions and a 3% mismatch in the third OP direction. Our films are a remarkable example of a phase transformation that can occur laterally and one where the transformation volume change can be accommodated fully by the free surface (by raising or depressing the film surface), which minimizes strain. We anticipate these pathways will be useful for designing fast-switching phase-change thin film devices that harness displacive phase transitions and the large contrast in optical and thermal properties between 3D-bonding and layered/2D-bonding.

**Temperature-induced phase transition of an as-grown rocksalt film**

We directly observe a temperature-induced transformation from 3D to 2D-bonding in MBE films by cooling Pb-rich films to cryogenic temperatures. Figure 10a shows the room temperature RSM of a $Pb_{0.58}Sn_{0.42}Se$ single-phase rocksalt film grown with a thin PbSe buffer layer. As this sample is cooled, a layered *Pnma* phase emerges (see also Fig. S7), and the film has a very strong layered film peak with a weak rocksalt secondary peak at −175 °C (Fig. 10a). This is maintained even when the film is brought back to room temperature (Fig. 10a). Although the rocksalt peak intensity is reduced, we do not see any shifts corresponding to an enlargement in the lattice constant from Pb-enrichment. The OP lattice constant of the orthorhombic phase extracted from the RSM after heating back up to room temperature (Fig. S8) also fits well with our trend of OP lattice constant vs alloy composition (Fig. 3c). This suggests the formation of a $Pb_{0.58}Sn_{0.42}Se$ layered phase. It is reasonable to expect a freezing of atom diffusion at these ambient and cryogenic temperatures and over short time scales involved in the experiment. Hence the transformation from rocksalt to layered-orthorhombic must be displacive.

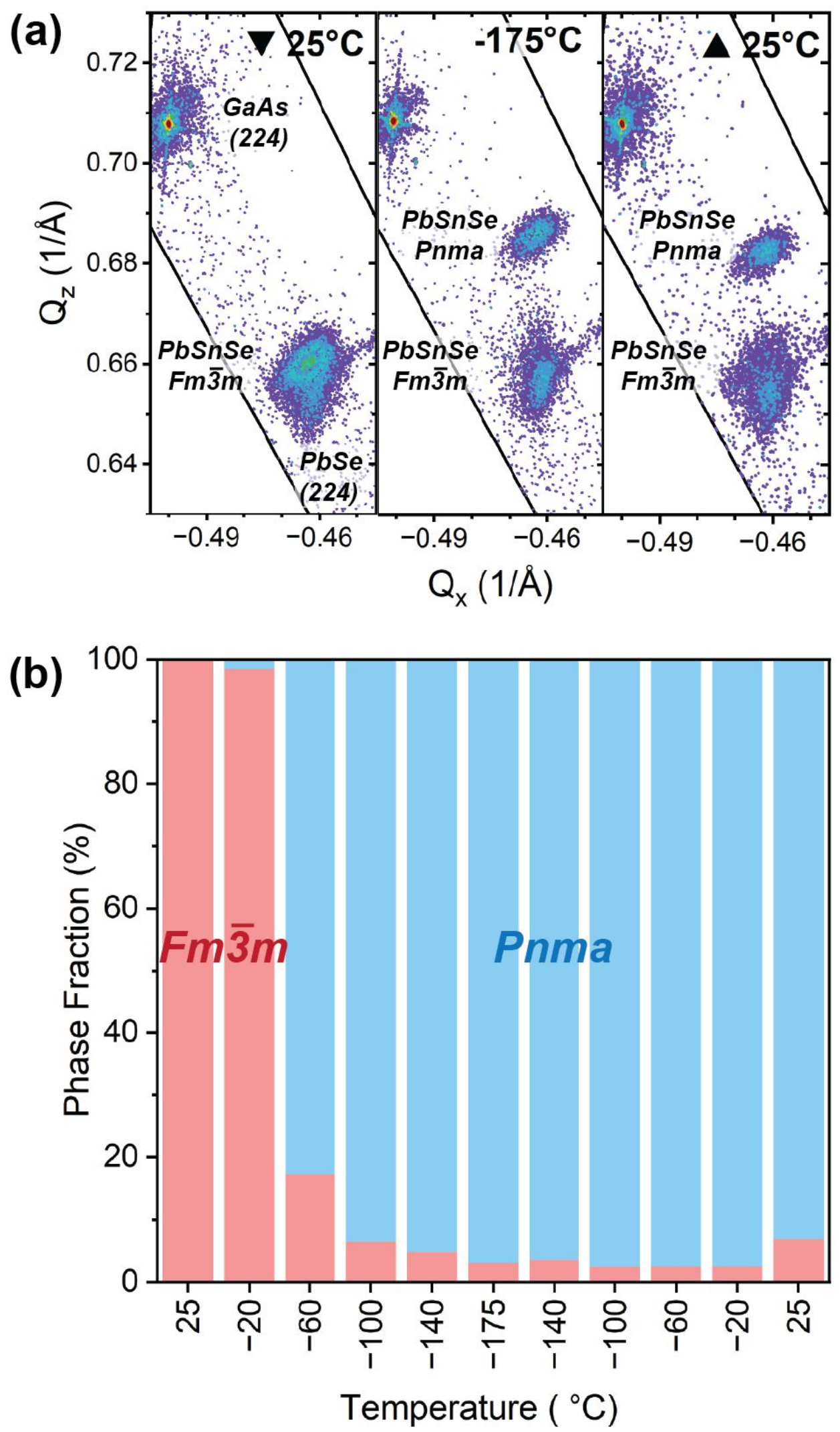

**Figure 10.** RSMs of an as-grown $X_{Sn}$=0.42 single-phase rocksalt PbSnSe film taken at (a) room temperature, after being cooled to -175 °C using liquid nitrogen, and after warming back up to room temperature, showing that the single-phase rocksalt sample becomes two-phase layered and rocksalt. (b) Phase fractions extracted from temperature-dependent RSM data, showing an abrupt transition to a mainly layered phase sample, with a slight recovery of the rocksalt phase when the sample is warmed back to room temperature.

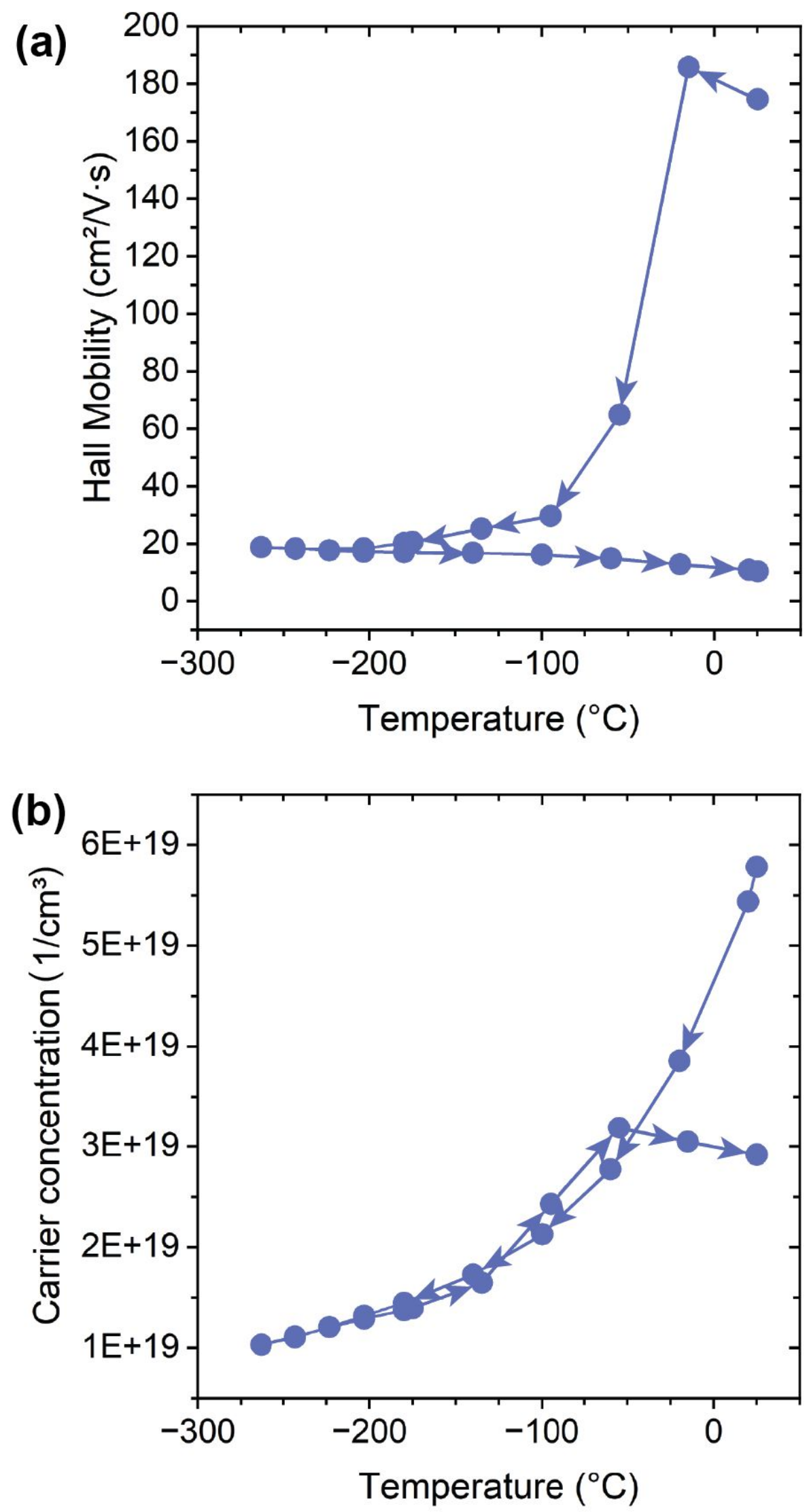

**Figure 11.** Temperature dependent (a) hall mobility and (b) carrier concentration of an as-grown $X_{Sn}$=0.42 single-phase rocksalt PbSnSe film, as it is cooled to -260 °C and heated back to room temperature.

Using known structure factor information of the rocksalt and layered structure, we calculate the phase fractions present at each measured temperature point (Fig. 10b). Indeed, the sample transforms sharply to a majority layered phase sample on cooling. On re-heating, the proportion of the rocksalt phase only begins increasing around room temperature (Fig. 10b). Hysteretic behavior is not unexpected in diffusionless transformations; Katase *et al*. and Nishimura *et al.* report similar behavior to their observed phase transitions in PbSnSe.[17,23] Thus we expect that if we heated this sample further, it would transition back to a single-phase rocksalt sample but we do not attempt this here due to the potential for Se loss from the sample at elevated temperatures.

Cryogenic Hall measurements also capture electronic property change with the temperature-induced phase-change. The Hall mobility of the as-grown rocksalt p-type film starts around 180 $cm^2/V\cdot s$ and abruptly drops with the phase transition on cooling to around 20 $cm^2/V\cdot s$ for the majority layered phase film (Fig. 11a), which is expected.[23,17] However, the carrier concentration does not have a similarly dramatic trend with phase-change. The carrier concentration decreases with cooling, and when heated to room temperature again, increases to about half the rocksalt phase value, remaining p-type in the range of $10^{19}/cm^3$ (Fig. 11b). While this is an expected carrier concentration for cubic PbSnSe due to a propensity for Sn vacancies in the rocksalt structure,[53,54] it is an unusually high carrier concentration for the resulting layered phase.[55] For reference, layered films grown without a PbSe buffer are p-type with hole concentrations only in the range of $10^{14}$-$10^{15}$ $/cm^3$ (Table S1). Unlike the rocksalt phase, layered PbSnSe is expected to be only lightly p-type due to different Sn vacancy energetics.[56] For reference, we measured a single-phase layered $Pb_{0.35}Sn_{0.65}Se$ film, with the same thickness and PbSe buffer layer, as n-type with a carrier concentration of $4.5 \times 10^{17}$ $/cm^3$. This overall n-type conduction is due to the overwhelming presence of Se vacancies in the PbSe buffer layer.[57] Thus, in comparing the as-grown layered film and the layered film resulting from a displacive phase transformation, both with a PbSe buffer layer, the carrier concentration of the phase-transformed film is orders of magnitude higher and of the opposite carrier type. We suspect that the very high hole concentration in the phase-transformed layered film is a result of the sample retaining vacancies from the rocksalt phase across the phase-change. Therefore, inducing a phase-change at low temperatures may be a feasible way to engineer metastable defect concentrations and achieve doping not otherwise possible without extrinsic impurities.

**Summary and conclusions**

We have shown methods to grow high quality epitaxial SnSe and PbSnSe alloys on technologically relevant GaAs substrates at low temperatures up to 300 °C. Importantly, this method allows us to directly grow metastable PbSnSe films. To explore the two closely related phases of the PbSe-SnSe materials system, rocksalt and layered, we have quantified how structural properties and active Raman modes change with alloy composition, and indeed see that lattice constants approach that of the rocksalt phase with increased alloying of PbSe into SnSe. Importantly, we observe that with epitaxial films grown by MBE, the phase space of as-grown films on the layered side is expanded from around $Pb_{0.25}Sn_{0.75}Se$ in the bulk to at least $Pb_{0.45}Sn_{0.55}Se$. Going forward, these high Pb-content *Pnma* alloys may provide a route to the high temperature *Cmcm* phase under more accessible conditions. Recent calculations suggest that the *Cmcm* phase has a significantly narrower bandgap than the *Pnma* phase with the potential also to have a direct electronic transition,[58] and could have application in moderate-temperature thermoelectrics and infrared optoelectronics.

We grew films close to $Pb_{0.50}Sn_{0.50}Se$ that have a microstructure consistent with a displacive phase transition between the rocksalt and layered phase rather than a conventional diffusive transformation. Using *in situ* temperature-dependent XRD measurements, as well as microscopy techniques such as STEM and EDS, we speculate that the samples were rocksalt at growth temperature and partially transform into the layered-phase on cool down. Microstructure characterization reveals two types of interfaces between the phases with well-defined orientation relationships and novel strain-relief mechanisms. The transformation between a 3D to 2D structure occurring laterally and with minimal strain is particularly exciting for heterostructure devices where optical, electronic, and thermal properties may be switched. Cryogenic Hall measurements show changes with phase transformation and reveal that this temperature induced transformation may enable highly doped layered phase PbSnSe films. There remain several challenges to develop this material system for phase-change devices. The temperature swing over which the samples change phase remains high, on the order of a hundred degrees Celsius, and thus heating or cooling the active volume limits the switching speed and energy cost. Although progress has been made recently to narrow this temperature swing in the related PbSnS materials system,[59] the kinetics of this phase transformation also requires more study to understand obstacles to complete phase transformation and reversibility. If direct light-field switching is feasible in SnSe, this thin film approach on GaAs provides new avenues to both tune the energetic landscape using alloys, and to couple these materials with existing photonic schemes.

## Methods

### PbSnSe thin film synthesis

The PbSnSe alloys in this study are grown by solid source MBE using PbSe and SnSe compound sources in standard effusion cells. Films were grown on epi-ready (001)-oriented semi-insulating substrates of GaAs with no intentional miscut. The oxide layer on the substrate was desorbed at 560 °C, as measured using a pyrometer in the presence of a Se overpressure from a separate valved cracker cell, for 10 minutes. RHEED pattern shows the 2 × 1 reconstruction, typical of a Se-terminated GaAs(001) surface.[27,60] Further surface preparation steps described in the Results section were used, after which approximately 300 nm of the desired PbSnSe alloy was grown by controlling the flux ratio between the PbSe and SnSe compound cells, with the SnSe beam equivalent pressure (BEP) fixed at $3\times10^{-7}$ Torr.

**Structural characterization**

The morphology of grown films was characterized using SEM on a TFS Apreo-S microscope at 5 kV. The out of plane orientation of the film, in-plane and out of-plane lattice constants, and film thickness were determined through XRD via symmetric 2θ-ω scans in a triple-axis configuration, RSMs, and X-ray reflection (XRR) data in a PANalytical X'Pert Pro system. Temperature-dependent RSM XRD data was collected on a PANalytical Empyrean X-ray diffractometer system using an Anton Paar domed cooling-heating stage. TEM foils were prepared on a FEI Helios NanoLab 600i DualBeam SEM/Focus Ion Beam system. TEM and STEM were taken on a FEI Tecnai G2 F20 X-TWIN TEM with a 200 kV electron beam, and atomic resolution STEM images were taken on a Thermo Fisher Spectra 300 with a 300 kV electron beam.

**Compositional analysis**

For elemental analysis, SEM-EDS was performed on a JEOL JSM-IT500HR environmental scanning electron microscope at 10 kV. We used the mα, lα, and lα transitions to identify the presence of Pb, Sn, and Se respectively. STEM-EDS data was collected on a Thermo Fisher Spectra 300 with a 300 kV electron beam. Data was quantified with a standard Cliff-Lorimer fit using default k-factors available in the TFS Velox software. XPS analysis was carried out with a Thermo Fisher Escalab $Xi^+$ instrument using a monochromatic Al $K_{alpha}$ anode (S1).

**Property characterization**

Raman spectra were collected on a HORIBA Scientific LabRAM HR Evolution spectrometer with an unpolarized 785 nm laser at 2.5% output power with a 1800 gr/m grating, and a 300 second accumulation. Hall measurements were conducted on a Lakeshore 8404 measurement tool. Room temperature measurements were conducted with an oven head, and *in situ* cryogenic measurements were conducted with a closed cycle refrigerator head, both using a 0.9 T field.

**Acknowledgements**

We gratefully acknowledge support via the NSF CAREER award under Grant No. DMR-2036520. We made use of the Stanford Nano Shared Facilities (SNSF) supported by the NSF under Award No. ECCS-2026822 for various materials characterization techniques. This research was conducted with government support awarded by the National Science Foundation Graduate Research Fellowship under Grant No. DGE-1656518, which funded P. Reddy. We are thankful for discussions with W. Nix and E. Hughes, and support for electron microscopy and interpretation from B. Haidet, P. Mukherjee, and A. Barnum.

**Supporting Information Available:** XPS data error analysis; XRD phase analysis; additional RSMs at room temperature and as a function of sample temperature; FWHM of alloy series rocking curves; RHEED comparison between epitaxial and polycrystalline films; SEM-EDS of two-phase film; additional high resolution HAADF STEM images of two-phase sample; lattice constant trends as a function of temperature for an as-grown rocksalt sample; Hall data from select films; composition summary of all PbSnSe film that are part of this study.

## Supporting Information

**Expanded stability of layered SnSe-PbSe alloys and evidence of displacive phase transformation from rocksalt in heteroepitaxial thin films**

Pooja D. Reddy[1*], Leland Nordin[1], Lillian Hughes[2], Anna-Katharina Preidl[1], Kunal Mukherjee[1†]

[1]Department of Materials Science and Engineering, Stanford University, Stanford, California 94306, USA

[2]Materials Department, University of California, Santa Barbara, California 93106, USA

* poojadr@stanford.edu
† kunalm@stanford.edu

**XPS data error explanation S1.**

The XPS analyses were carried out with a ThermoFisher Escalab $Xi^+$ instrument using a monochromatic Al $K_{alpha}$ anode (E = 1486.7 eV). Survey scan analyses were carried out with an analysis area of 400 x 400 μm and a pass energy of 100 eV. A depth profile was formed by taking sequential scans after 15-25 rounds of sputtering for 20 seconds with an $Ar^+$ ion beam (2 kV). The Sn3d3, Pb4d5, and Se3d peaks were fit using AvantageXPS software with Smart backgrounds applied, and atomic fractions quantified at each depth to calculate average elemental composition throughout the film. Figure S1a shows an example spectrum from the $Pb_{0.31}Sn_{0.69}Se$ sample with the selected peaks highlighted. Sn3d3 and Pb4d5 are chosen not only because they give a large, isolated signal but also because they are close in binding energy and will exhibit a similar mean free path, which helps minimize error in quantification. The variation in composition throughout the film thickness is minimal, as can be seen in the depth profile for the SnSe sample (Fig. S1b).

Compared to STEM-EDS, XRD, and Raman results, XPS seems to overestimate Sn composition. When running standards in our instrument, we observe ± 4% relative variation and therefore report the compositions only up to two significant digits. Because of the surface sensitivity of XPS, a large source of error is the in-depth distribution of quantified elements. In our case this concern is minimized because we calculate average composition throughout the entire film thickness. Preferential etching of lighter elements could be another error source but is unlikely to have a significant impact for Pb and Sn as they are both much heavier than Ar. Lastly, we do not observe significant changes in the quantifications with adjustments to the integration bounds and therefore do not attribute peak fitting as a major source of error in our estimates of film composition.

**Discerning which orthorhombic phase is present in films S2.**

SnSe has two orthorhombic phases, the *Pnma* phase, which is stable at lower temperatures, and the *Cmcm* phase, which is stable above temperatures of about 600 °C. While both are layered orthorhombic, the *Cmcm* structure has in-plane lattice constants that are essentially equal, around 4.3 Å for pure SnSe. Some of the samples grown with higher Pb compositions have layered phases with essentially identical in-plane lattice constants, such that the sample RSMs about the 224 GaAs peak showed only one layered film peak (Fig 4), rather than the two we expect for the two ways a layered unit cell can arrange on a cubic substrate (Fig 1d). Since the in-plane lattice constants are equal, one might think that the layered film phase corresponds to the *Cmcm* phase. We do see that as the Pb composition of layered PbSnSe increases, the in-plane lattice constants approach each other (Fig 3c). Therefore, the film could also just be the *Pnma* phase whose lattice constants changed with composition such that they now seem equal.

The *Cmcm* space group has more symmetry than the *Pnma* space group. Thus, we expect that the XRD pattern of the *Cmcm* phase would have less allowed peaks than the *Pnma* phase. We found that the (811) film peak is expected to be present in the *Pnma* phase, and is a forbidden peak for the *Cmcm* phase. So, to discern whether layered phases with nearly identical in-plane lattice constants are *Cmcm* or *Pnma*, we take RSMs of these samples around where we expect the (811) peak to be. Figure S2 shows such RSMs for pure SnSe grown at 300 °C, and two samples where the layered phase has nearly identical in-plane lattice constants: a two-phase sample grown at 300 °C where the layered and rocksalt phase have compositions of $X_{Sn}$=0.53 and $X_{Sn}$=0.49 respectively, and a single-phase layered sample with a composition of $X_{Sn}$=0.67, grown at 210 °C. As expected, the SnSe film which is the *Pnma* phase, has the (811) peak present in the RSM about the 024 GaAs peak. And the other two samples also show the (811) peak. This means that the samples grown which seem to have nearly identical in-plane lattice constants are actually all in the *Pnma* phase.

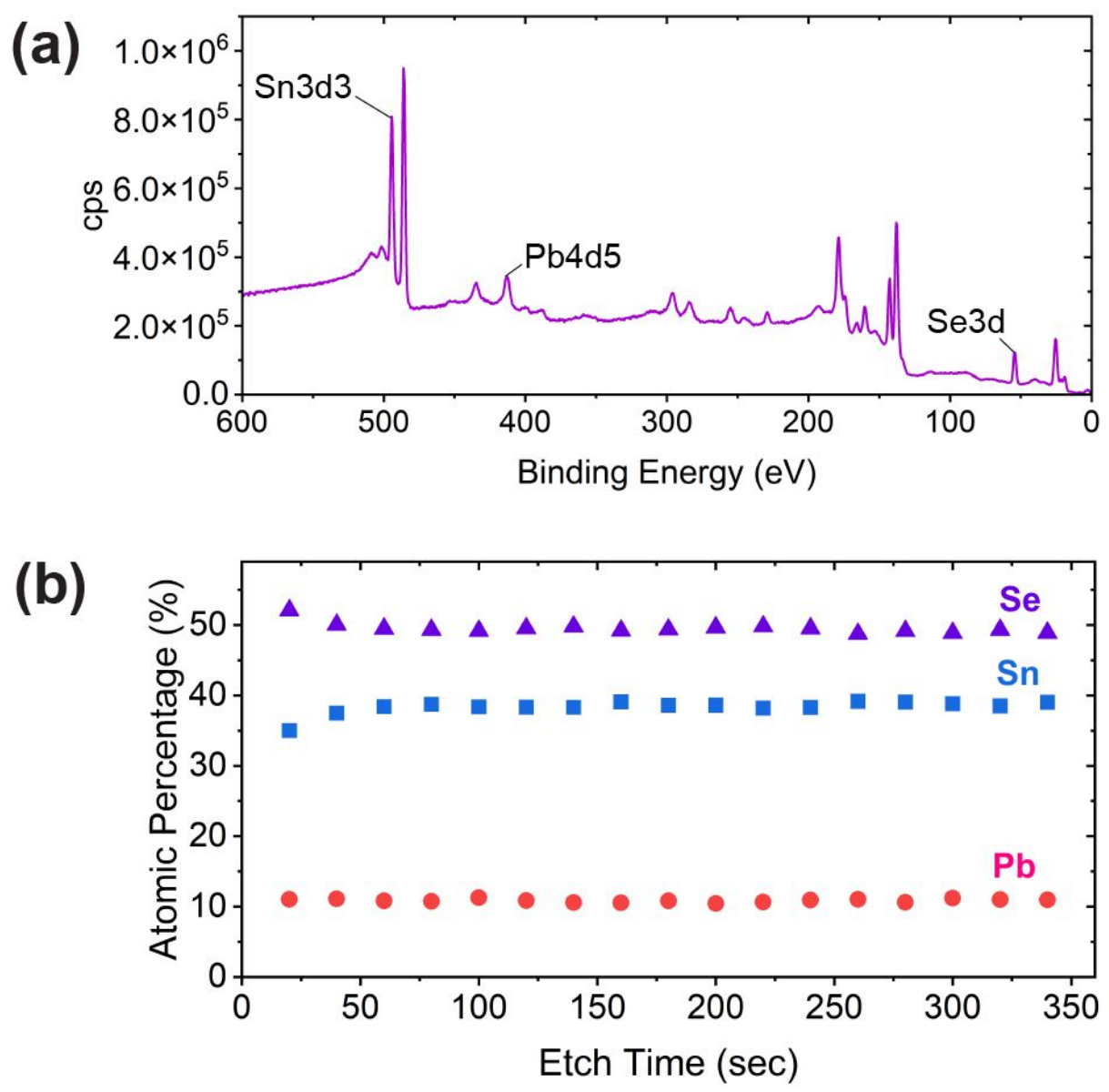

**Figure S1.** (a) Example XPS survey scan (100 eV pass energy) of the layered phase $Pb_{0.31}Sn_{0.69}Se$ sample grown at 300 °C, with quantified peaks identified.(b) Depth profile of the quantified Sn3d3, Pb4d5, and Se3d peaks show That composition is consistent throughout the epilayer thickness.

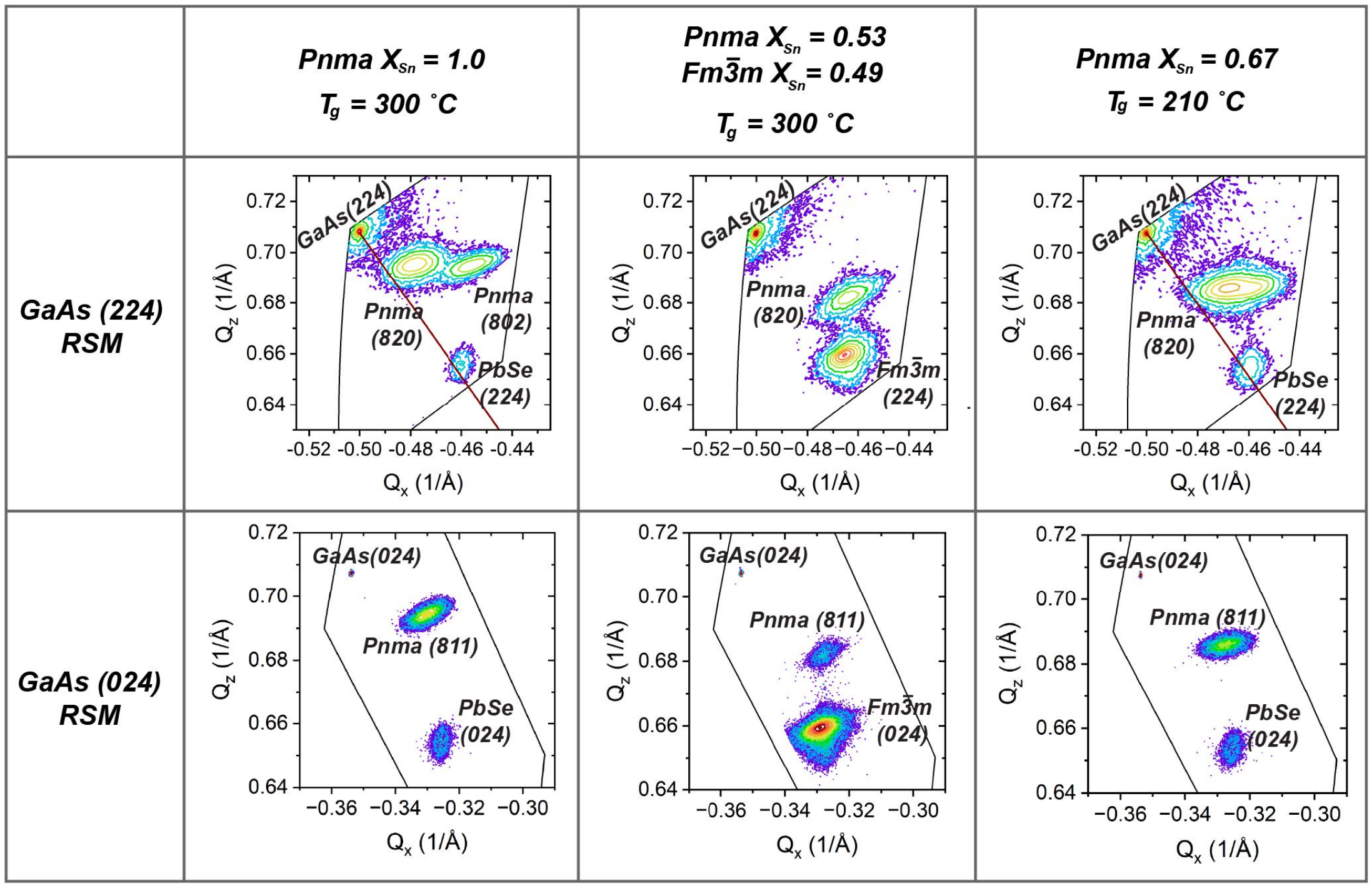

**Figure S2.** RSMs around the (224) and (024) peaks of GaAs for select films, with present phase peaks labelled.

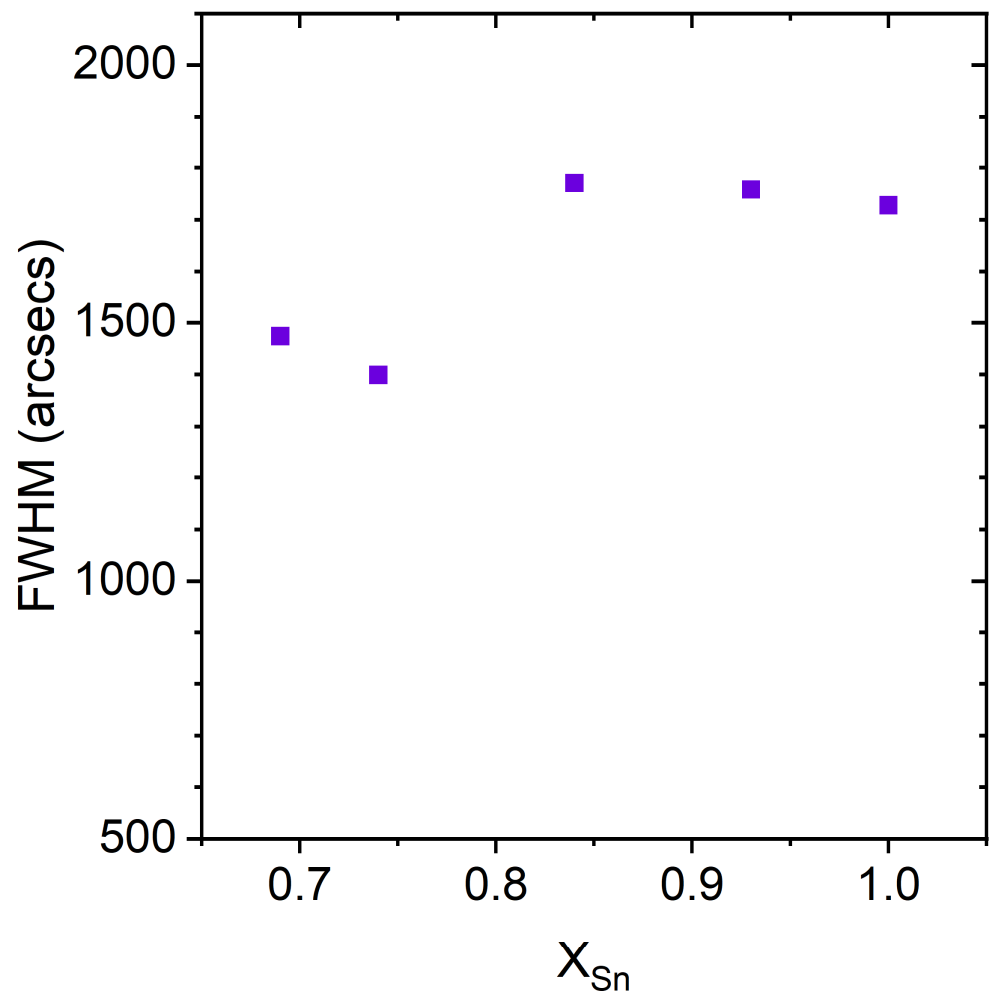

**Figure S3.** The FWHM of the double-axis rocking curves of the (800) film peak reflections, of the layered phase $Pb_{1-x}Sn_xSe$ series grown at $T_g$=300 °C.

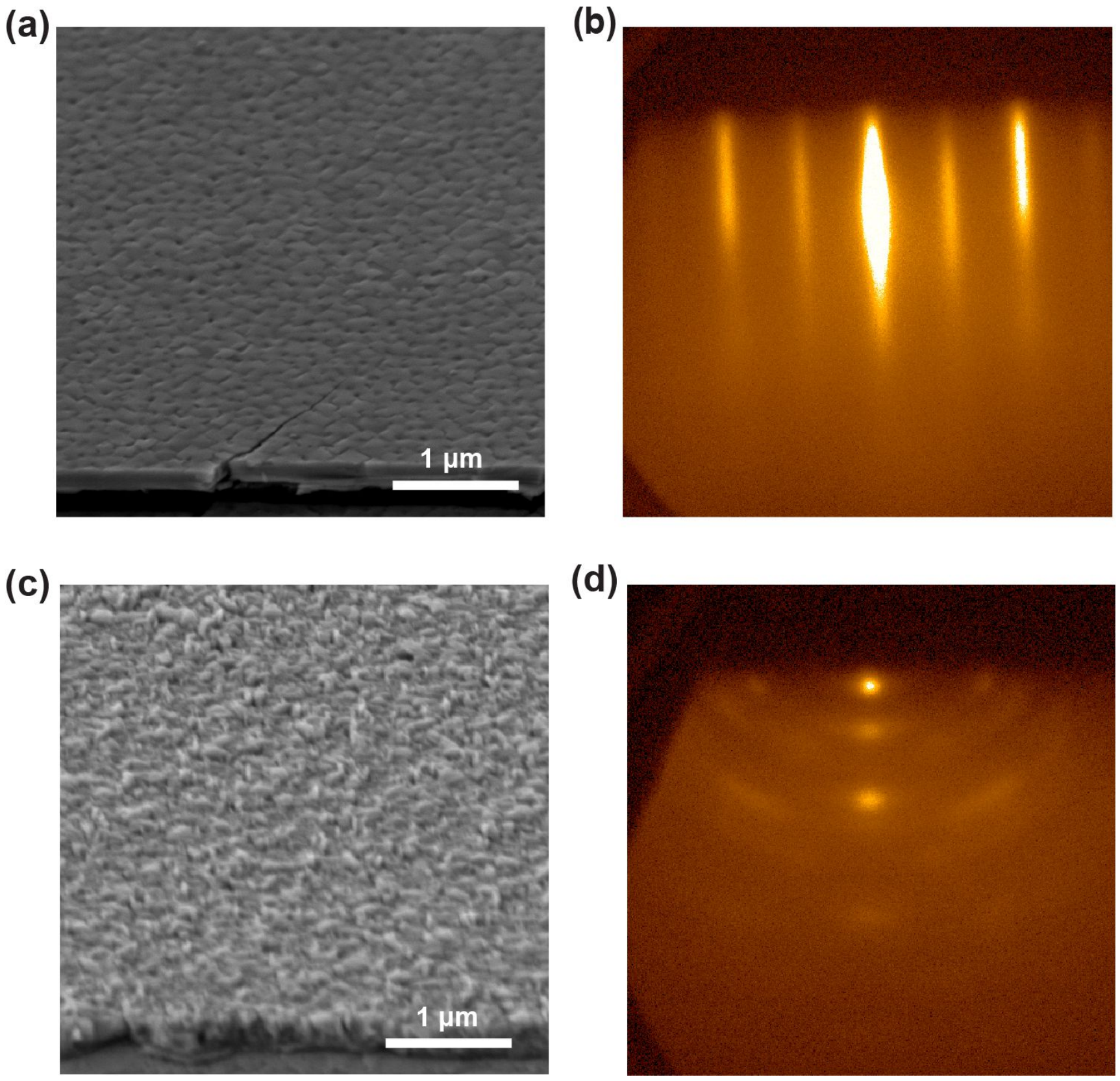

**Figure S4.** An epitaxial film grown with 1:1 BEP ratio of SnSe:PbSe at 270 °C, (a) SEM image of the 45° mounted surface and (b) RHEED image of film after growth. In comparison, a polycrystalline film grown with 1:1 BEP ratio of SnSe:PbSe at 150 °C, (c) SEM image of the 45° mounted surface and (d) RHEED image of film after growth.

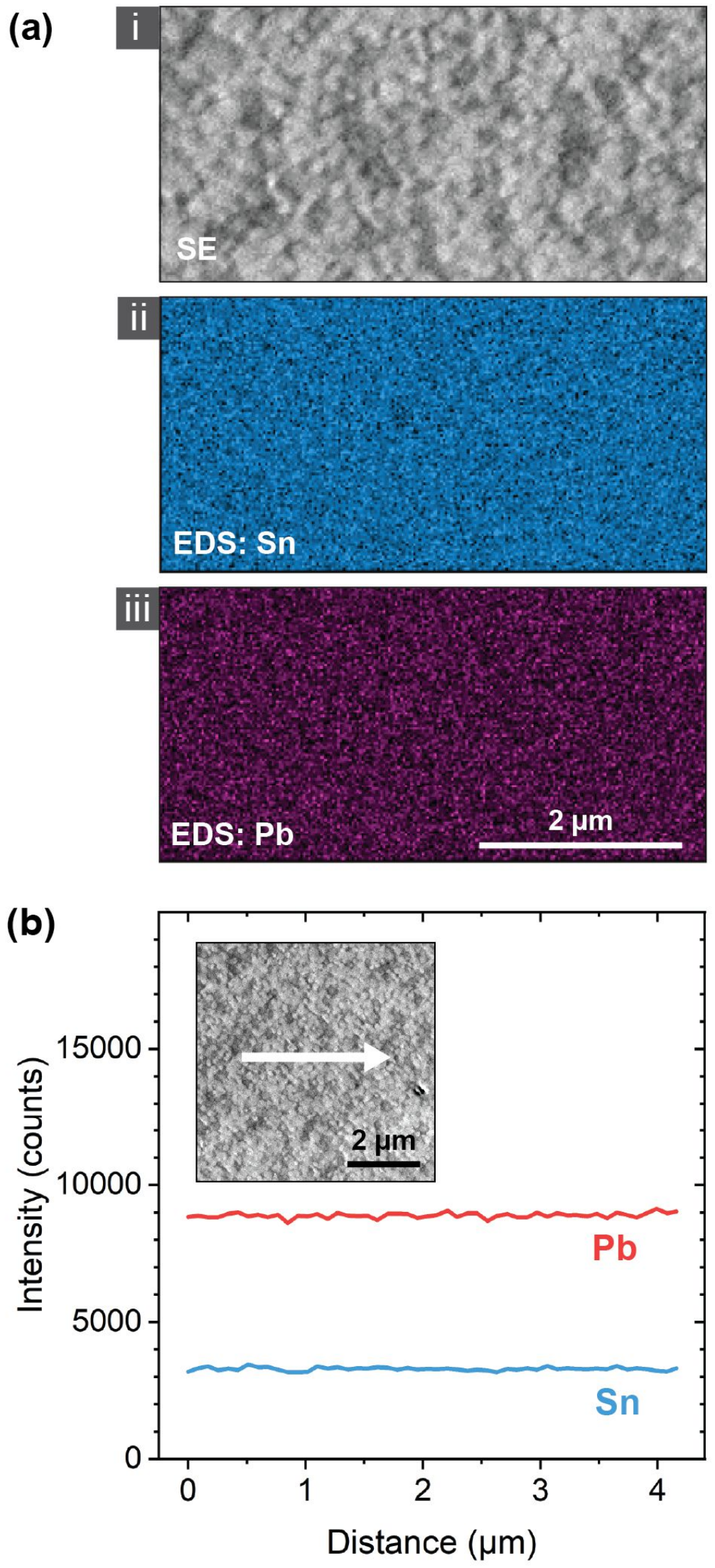

**Figure S5.** Standardless SEM-EDS data of a PbSnSe alloy grown at 300 °C with layered and rocksalt phase compositions of approximately $X_{Sn}$=0.53, and $X_{Sn}$=0.49 respectively. (a) Film surface (i) in secondary electron SEM mode (ii) showing Sn EDS map (iii) showing Pb EDS map. (b) EDS line scan of the area of the sample marked in the inset SEM image. Different element transitions have different detection sensitivities, so while the alloy composition is about 1:1, we do not expect the intensity of Pb and Sn signal to be equal in SEM-EDS.

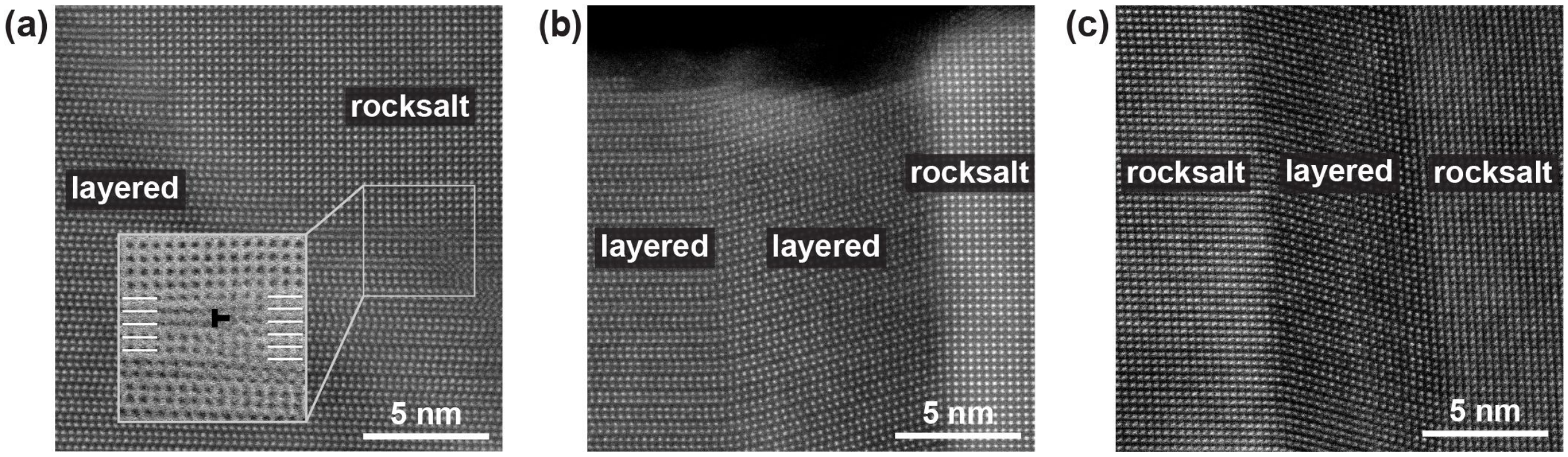

**Figure S6.** HAADF STEM images of two-phase sample grown at 300 °C with layered and rocksalt phase compositions of approximately $X_{Sn}$=0.53, and $X_{Sn}$=0.49 respectively. a) Image of a curved boundary. The inset bright field image of the marked area highlights a dislocation in the sample. (b) Image of a vertical boundary at the sample surface, showing the layered phases are depressed compared to the rocksalt phase. (c) Image of vertical grain boundaries between rocksalt, tilted layered, and rocksalt grains.

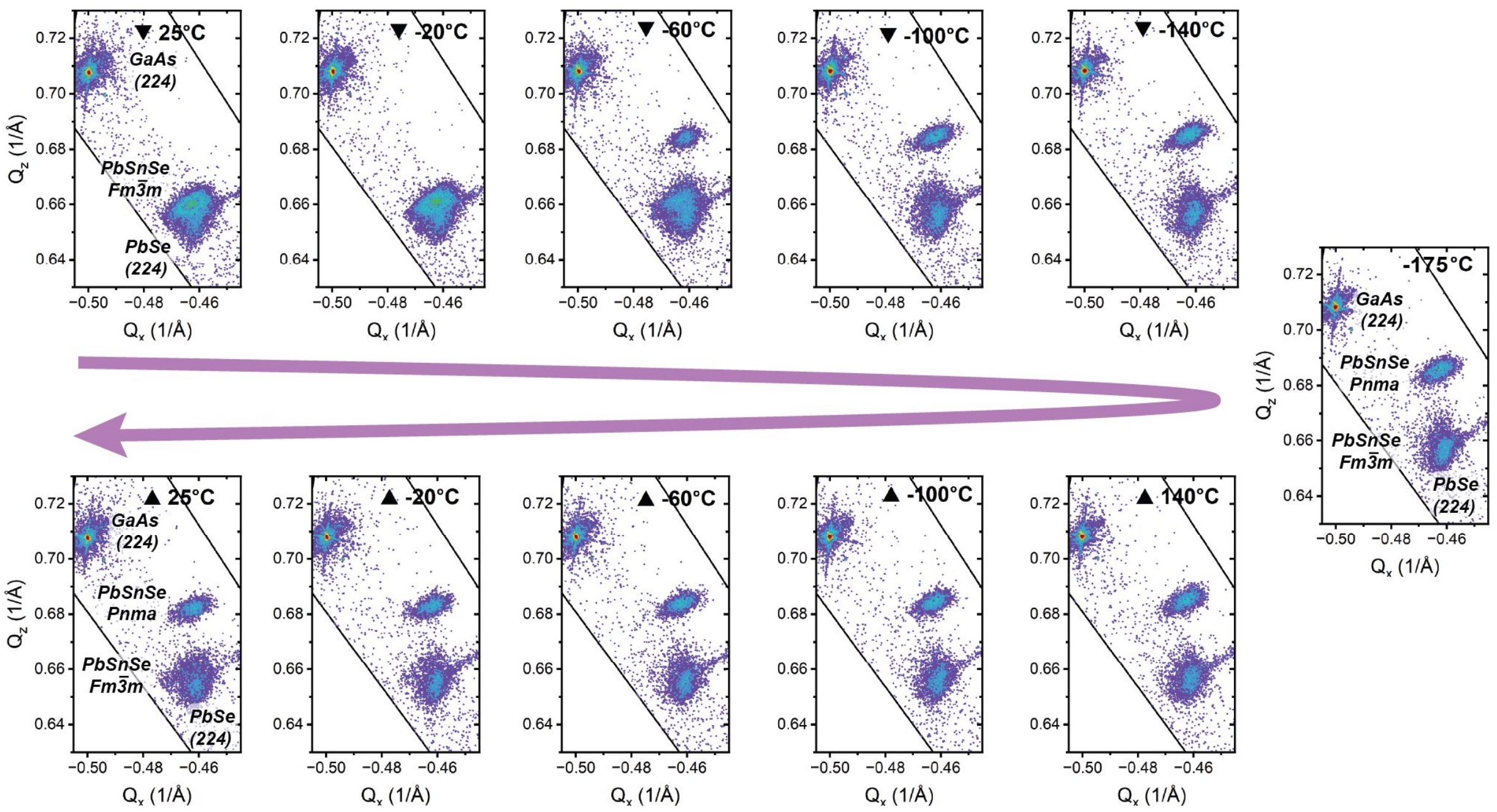

**Figure S7.** RSMs taken at 40 °C intervals of an as grown $X_{Sn}$=0.42 single phase cubic PbSnSe film as it is temperature cycled starting at room temperature, cooling to -175 °C using liquid nitrogen, and warming up back up to room temperature. An orthorhombic phase appears on the cool down at around -60 °C.

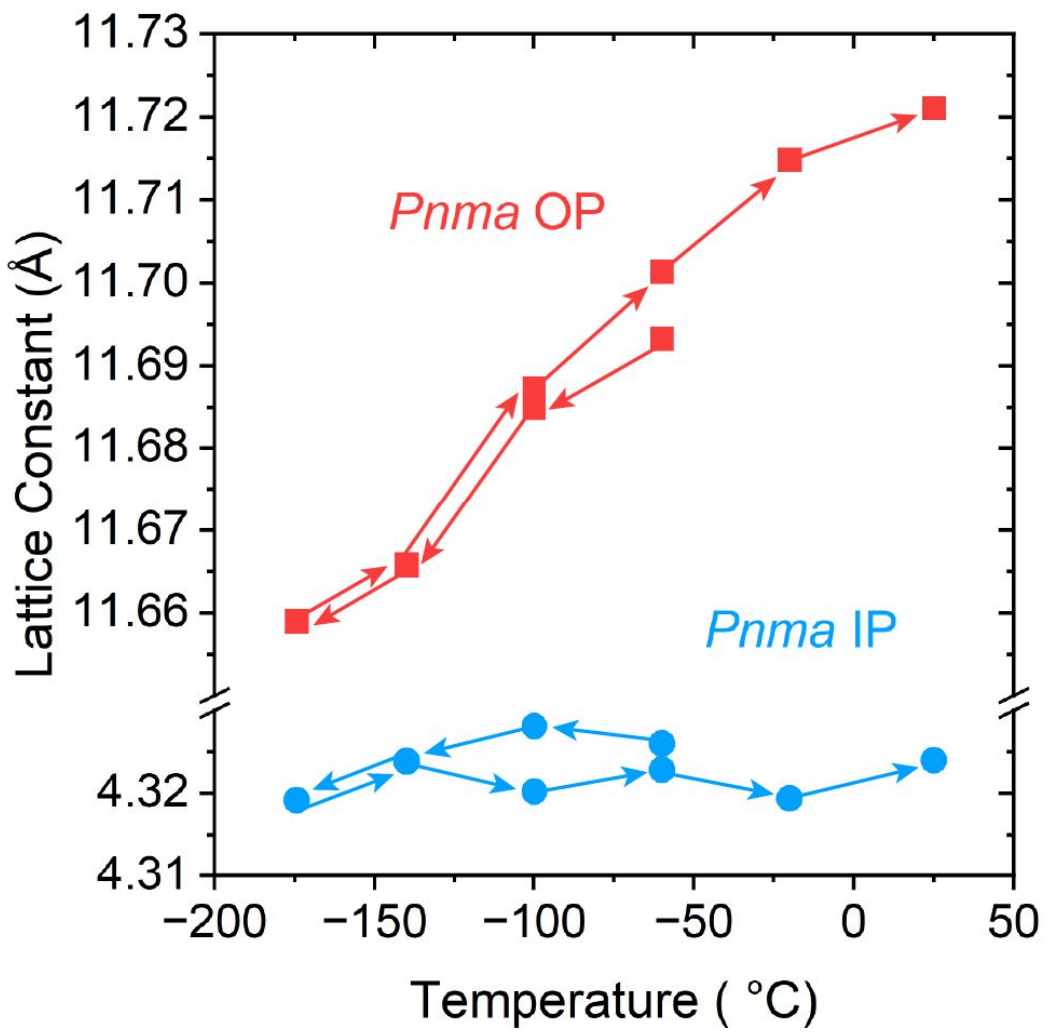

**Figure S8.** Lattice constant data of an as grown $X_{Sn}$=0.42 single phase cubic PbSnSe film extracted from temperature cycling RSM data, including the trend of the OP layered phase lattice constant, and the [001] IP direction of the layered phase.

**Table S1.** Select electronic properties of SnSe films shown in Figure 1a of the main manuscript, grown with different surface preparation steps. Approximate thickness values were used for calculations and changing these within reason did not affect order of magnitude of the results.

| Surface Preparation | Carrier Type | Carrier Concentration (1/cm³) | Hall Mobility (cm²/V·s) |
|---|---|---|---|
| SnSe /GaAs | p | $4 \times 10^{15}$ | 8 |
| SnSe/Se dose/ GaAs | p | $1 \times 10^{15}$ | 9 |
| SnSe/PbSe dose/ GaAs | p | $1 \times 10^{14}$ | 48 |
| SnSe/PbSe/PbSe dose/GaAs | n | $4 \times 10^{16}$ | 47 |

**Table S2.** Summary of PbSnSe alloys grown as part of this study. Compositions are found via XRD data in conjunction with published bulk data sets, as described in the main manuscript.

| Growth Temperature (°C ) | Phases Present | Composition ($X_{Sn}$) | |
|---|---|---|---|
| | | *Pnma* | $Fm\bar{3}m$ |
| 300 | *Pnma* | 1.00 | -- |
| 300 | *Pnma* | 0.93 | -- |
| 300 | *Pnma* | 0.84 | -- |
| 300 | *Pnma* | 0.74 | -- |
| 300 | *Pnma* | 0.69 | -- |
| 300 | *Pnma* + $Fm\bar{3}m$ | 0.53 | 0.49 |
| 300 | *Pnma* + $Fm\bar{3}m$ | 0.52 | 0.49 |
| 270 | *Pnma* + $Fm\bar{3}m$ | 0.52 | 0.51 |
| 255 | *Pnma* | 0.71 | -- |
| 250 | *Pnma* | 0.65 | -- |
| 250 | $Fm\bar{3}m$ | -- | 0.42 |
| 240 | *Pnma* | 0.71 | -- |
| 210 | *Pnma* | 0.67 | -- |
| 195 | *Pnma* + $Fm\bar{3}m$ | 0.53 | 0.47 |
| 180 | *Pnma* | 0.74 | -- |
| 180 | *Pnma* | 0.66 | -- |
| 180 | $Fm\bar{3}m$ | -- | 0.44 |
| 165 | *Pnma* | 0.55 | -- |